\documentclass[twocolumn]{aastex63}

\usepackage{textcomp}
\usepackage{comment}
\usepackage{physics}
\usepackage{lineno}

\submitjournal{ApJ}

\shorttitle{IMBH candidates from ZTF and \textit{WISE}}
\shortauthors{Ward et al.}

\graphicspath{{./}{figures/}}

\begin{document}
\title{Variability--selected intermediate mass black hole candidates in dwarf galaxies from ZTF and \textit{WISE}}
\correspondingauthor{Charlotte Ward}
\email{charlotteward@astro.umd.edu}
\author[0000-0002-4557-6682]{Charlotte Ward}
\affil{Department of Astronomy, University of Maryland, College Park, MD  20742, USA}
\affiliation{Lawrence Berkeley National Laboratory, 1 Cyclotron Road, Berkeley, CA 94720, USA}
\author[0000-0003-3703-5154]{Suvi Gezari}
\affil{Space Telescope Science Institute, 3700 San Martin Dr., Baltimore, MD 21218, USA}
\affil{Department of Astronomy, University of Maryland, College Park, MD  20742, USA} 
\author[0000-0002-3389-0586]{Peter Nugent}
\affiliation{Lawrence Berkeley National Laboratory, 1 Cyclotron Road, Berkeley, CA 94720, USA}
\affiliation{Department of Astronomy, University of California, Berkeley, Berkeley, CA 94720, USA}
\author[0000-0001-8018-5348]{Eric C. Bellm}
\affiliation{DIRAC Institute, Department of Astronomy, University of Washington, 3910 15th Avenue NE, Seattle, WA 98195, USA}
\author{Richard Dekany}
\affiliation{Caltech Optical Observatories, California Institute of Technology, Pasadena, CA 91125, USA}
\author{Andrew Drake}
\affiliation{Division of Physics, Mathematics, and Astronomy, California Institute of Technology, Pasadena, CA 91125, USA}
\author[0000-0001-5060-8733]{Dmitry A. Duev}
\affiliation{Division of Physics, Mathematics, and Astronomy, California Institute of Technology, Pasadena, CA 91125, USA}
\author[0000-0002-3168-0139]{Matthew J. Graham}
\affiliation{Division of Physics, Mathematics, and Astronomy, California Institute of Technology, Pasadena, CA 91125, USA}
\author[0000-0002-5619-4938]{Mansi M. Kasliwal}
\affil{Division of Physics, Mathematics, and Astronomy, California Institute of Technology, Pasadena, CA 91125, USA}
\author[0000-0002-7252-3877]{Erik C. Kool}
\affiliation{The Oskar Klein Centre, Department of Astronomy, Stockholm University, AlbaNova, SE-10691, Stockholm, Sweden}
\author[0000-0002-8532-9395]{Frank J. Masci}
\affiliation{IPAC, California Institute of Technology, 1200 E. California Blvd, Pasadena, CA 91125, USA}
\author{Reed L. Riddle}
\affiliation{Division of Physics, Mathematics, and Astronomy, California Institute of Technology, Pasadena, CA 91125, USA}
\begin{abstract}
While it is difficult to observe the first black hole seeds in the early Universe, we can study intermediate mass black holes (IMBHs) in local dwarf galaxies for clues about their origins. In this paper we present a sample of variability--selected AGN in dwarf galaxies using optical photometry from the Zwicky Transient Facility (ZTF) and forward--modeled mid--IR photometry of time--resolved Wide--field Infrared Survey Explorer ({\it WISE}) coadded images. We found that 44 out of 25,714 dwarf galaxies had optically variable AGN candidates, and 148 out of 79,879 dwarf galaxies had mid--IR variable AGN candidates, corresponding to active fractions of $0.17\pm0.03$\% and $0.19\pm0.02$\% respectively. We found that spectroscopic approaches to AGN identification would have missed 81\% of our ZTF IMBH candidates and 69\% of our {\it WISE} IMBH candidates. Only $9$ candidates have been detected previously in radio, X-ray, and variability  searches for dwarf galaxy AGN. The ZTF and {\it WISE} dwarf galaxy AGN with broad Balmer lines have virial masses down to $10^{5.5}M_\odot$ and for the rest of the sample, BH masses predicted from host galaxy mass range between $10^{5.2}M_\odot<M_{\text{BH}}<10^{7.3}M_\odot$. We found that only 5 of 152 previously reported variability--selected AGN candidates from the Palomar Transient Factory in common with our parent sample were variable in ZTF. We also determined a nuclear supernova fraction of $0.05\pm0.01$\% year$^{-1}$ for dwarf galaxies in ZTF. Our ZTF and {\it WISE} IMBH candidates show the promise of variability searches for the discovery of otherwise hidden low mass AGN.
\vspace{1.5cm}
\end{abstract}

\section{Introduction}
It is challenging to determine how the very first massive black holes formed because they have grown over time as galaxies merge and their black holes accrete \citep{Volonteri2010Quasi-starsHoles}. High redshift black hole (BH) seeds are very difficult to detect due to their low luminosities \citep{Volonteri2016InferencesUniverse} so we can instead constrain models of BH seed formation by studying the least massive black holes in local galaxies \citep{Reines2016ObservationalSeeds}. These low mass analogs to black hole seeds are called intermediate mass black holes (IMBHs) and have masses in the range $100<M_{BH}<10^6 M_{\odot}$. Dwarf galaxies of stellar mass $M_* < 10^{9.75} M_{\odot}$ are the best place to find low mass black holes because galaxy mass and black hole mass are correlated \citep[e.g.][]{Kormendy2013CoevolutionGalaxies,Woo2013DoRelations}. Supernova--driven stunting of black hole growth in dwarf galaxies also makes these black holes comparable to early BH seeds \citep{Habouzit2017BlossomsFeedback, Angles-Alcazar2017BlackNuclei}.

The low mass end of galaxy population relations such as the black hole occupation fraction and the slope and scatter of the relationship between black hole mass and the stellar velocity dispersion of the host bulge ($M_{\mathrm{BH}}-\sigma_{*}$, \citet{Ferrarese2000AGalaxies}) will vary depending on the formation mechanism of early black hole seeds \cite[see][for a review]{Greene2020Intermediate-MassHoles}. For example, Pop III stars will produce a population of undermassive BHs in low redshift galaxies with stellar dispersion $\sigma_*<100$ km/s while direct collapse mechanisms will produce heavier black holes, resulting in a flattening of the $M_{\text{BH}}-\sigma$ relation around masses of $10^5 M_\odot\ $ \citep{Volonteri2009JourneyUniverse}. Although the potential for low mass black hole populations to constrain BH seed formation histories may be limited by uncertainties in accretion prescriptions \citep{Ricarte2018TheSeeds} it nonetheless motivates the discovery of a large population of black holes in low mass galaxies. A recent effort by \citet{Baldassare2020PopulatingRelation} to fill in the low mass end of the $M_{\mathrm{BH}}-\sigma_{*}$ relation doubled the number of black holes with measured virial masses and stellar velocity dispersions in dwarf galaxies of $M_* < 3\times 10^9 M_{\odot}$ to 15 and the results were in agreement with an extrapolation of the linear relationship observed for high mass galaxies. However, larger numbers of low mass AGN, particularly in galaxies of mass $M_*<3\times 10^8$, are needed to more fully constrain seed models. 

The fraction of IMBHs in dwarf galaxies which are wandering in their galaxy haloes, rather than occupying the nucleus, may also constrain BH seed formation mechanisms. The wandering fraction is dependent on galaxy merger history but will be substantially higher if massive black holes were produced by gravitational runaway of massive star remnants \citep{Miller2002ProductionClusters,PortegiesZwart2002TheClusters} due to the high frequency of IMBH ejection during these interactions \citep{Volonteri2005DynamicalUniverse,HolleyBockelmann2008GravitationalHoles}. \citet{Ricarte2021OriginsHoles} found that a substantial population of wandering black holes exist in the \texttt{ROMULUS} cosmological simulations and make up 10\% of black hole mass in the local universe.  Studies of IMBHs of mass $3.8<\log M_{\text{BH}}(M_\odot\ )<7.0$ in cosmological zoom--in simulations suggest that 50\% of IMBHs in dwarf galaxies are wandering within 7 kpc of the galaxy center due to historical galaxy mergers \citep{Bellovary2018MultimessengerGalaxies,Bellovary2021TheGalaxies}. This is supported by radio observations of dwarf galaxy AGN \citep{Reines2019AObservations}. Observational constraints on the wandering fraction of IMBHs in dwarf galaxies will help to test the accuracy of cosmological merger simulations, the effects of gravitational wave recoil on black holes in low mass galaxies and the feasibility of the gravitational runaway black hole seed formation mechanism.

A major challenge for both the search for IMBHs and for estimating the occupation and wandering fractions is that the predicted accretion rates are very low, particularly for non--nuclear AGN. Black hole accretion may be bimodal in its efficiency, causing low mass black holes $<10^5 M_\odot\ $to grow more slowly \citep{Pacucci2018GlimmeringFunction}. \citet {Bellovary2018MultimessengerGalaxies} found that most simulated IMBHs reach a maximum bolometric luminosity of $\log L_{\text{bol}} \text{(erg/s)}<41$ and are therefore very difficult to detect. \citet{Pacucci2021TheGalaxies} used a theoretical model based on galaxy mass and the angular momentum available in nuclear regions to estimate an active fraction of 5--22\% of AGN in dwarf galaxies which increases with host mass. The level of activity likely depends on the merger history of the dwarf galaxy. \citet{Kristensen2021MergerIllustrisTNG} find that inactive dwarf galaxies in the \texttt{Illustris} simulations tend to have been residing in dense environments for long times, while active galaxies had commonly been in a recent ($\leq$ 4 Gyr) minor merger. 

Discovery of AGN activity in dwarf galaxies will also help us to test predictions on the importance of AGN feedback in the low mass regime. While \citet{Geha2012AGalaxies} found that dwarf galaxies in SDSS with no active star formation are extremely rare and that more massive neighboring galaxies were the cause of star formation quenching, \citet{Penny2018SDSS-IVGalaxies} and \citet{Dickey2019AGNGalaxies} have found evidence for internal AGN--driven quenching in a small number of dwarf galaxies. Simulations suggest that dwarf galaxies hosting overmassive black holes have lower central stellar mass density, lower H I gas content and lower star formation rates than dwarf galaxies with undermassive counterparts, suggesting that internal feedback in dwarf galaxies can quench star formation only for higher mass black holes \citep{Sharma2019BlackGalaxies}. Simulations by \citet{Koudmani2019FastGalaxies} found that AGN outflows in dwarf galaxies only have a small effect on regulating global star formation rates compared to supernovae and sustained high--luminosity AGN with isotropic winds. 

Detailed observational studies of small dwarf galaxy samples have painted a different picture. GMOS IFU observations of 8 dwarf galaxies by \citet{Liu2020IntegralAGNs} discovered the presence of high velocity outflows in 7 out of 8, with some outflows capable of expelling a portion of outflowing material from the galaxy and enriching the surrounding circumgalactic medium. Coronal line emission inconsistent with shocks was also detected from 5 of these objects \citet{Bohn2021Near-InfraredOutflows}. Larger samples of AGN in dwarf galaxies will allow for further searches for signatures of AGN feedback such as high velocity and large scale outflows.

A range of approaches have been used to obtain samples of AGN candidates in dwarf galaxies. Some studies have taken a spectroscopic approach by looking for the emission line signatures of AGN in low mass galaxies. \citet{Reines2013DWARFHOLES} found that 151 dwarf galaxies with masses $10^{8.5}M_\odot\ < M_* < 10^{9.5} M_\odot\ $amongst a sample of 25,000 had [O\,{\sc iii}] $\lambda$5007/H$\beta$ and [N\,{\sc ii}] $\lambda$6550/H$\alpha$ narrow emission line ratios indicative of AGN activity. \citet{Mezcua2020HiddenAGN} found a sample of 37 dwarf galaxies with `hidden' AGN ionization lines by spatially resolving emission from star forming regions and nuclear AGN within each dwarf galaxy using MaNGA integral field unit spectroscopy. \citet{Molina2021AEvents} used [Fe\,X] $\lambda$6374 coronal line emission as a signature of AGN accretion to find 81 dwarf galaxies with possible IMBHs. 

Multi--wavelength approaches have also been successful in the identification of AGN in dwarf galaxies. \citet{Mezcua2018Intermediate-massSurvey} identified 40 AGN in dwarf galaxies with stellar masses $10^{7}M_\odot\ < M_* < 3\times 10^{9} M_\odot\ $ via their X-ray emission in the \textit{Chandra} COSMOS--Legacy Survey, finding an AGN fraction of $\sim$0.4\% for redshifts less than 0.3. \citet{Reines2019AObservations} found that 39 of 111 dwarf galaxies had compact radio sources at 8--12 GHz with the Karl G. Jansky Very Large Array (VLA) and determined that 13 of these could confidently be classified as AGN. They also found that 10 of the 13 radio AGN detected were spatially offset from their optical galaxy nuclei. 

A new approach to the detection of IMBH candidates has been to search for AGN--like stochastic variability from low mass galaxies as a signature of the presence of a central BH. \citet{Baldassare2018} found 135 galaxies with AGN--like variability on yearly timescales when constructing light curves of $\sim 28000$ galaxies with SDSS spectra, including 12 from dwarf galaxies with stellar masses $M_* < 3\times 10^{9} M_\odot\ $. They therefore estimated a variability fraction of 0.1\% for AGN in dwarf galaxies. A similar study in 2020 used light curves from the Palomar Transient Factory \citep{Law2009TheResults} to search for variable AGN in 35,000 galaxies with stellar mass $M_* < 10^{10} M_\odot\ $, identifying variability in 102 galaxies with masses $M_* < 3\times 10^{9} M_\odot\ $ \citep{Baldassare2020AFactory}.

As some low mass AGN vary on hourly timescales, very high cadence surveys have proved effective for the discovery of variability from dwarf galaxy AGN. For example, \citet{Burke2020OpticalSatellite} produced a 30 minute cadence, one month long light curve of the AGN in the archetypal dwarf galaxy NGC 4395 using data from the Transiting Exoplanet Survey Satellite (TESS). The $\sim10^5 M_{\odot}$ black hole was variable with a characteristic timescale of $1.4^{+1.9}_{-0.5}$ days. \citet{Martinez-Palomera2020IntroducingSurvey} used data from the HiTS imaging survey, which undertook one week of high cadence (4 times per night) and high coverage (120--150 deg$^2$) observations with the Dark Energy Camera each year for three years, to search for short timescale IMBH variability. They identified 500 galaxies with hourly, small amplitude variation in their week--long light curves. They estimated that 4\% of dwarf galaxies contained an IMBH based on their results.  \citet{Shaya2015ACTIVEMISSION} monitored $\sim$500 galaxies with the Kepler telescope \citep{Borucki2010KeplerResults} and found that while 4\% showed bright AGN activity, many other galaxies exhibited faint (down to 0.001 magnitude) variability which may also have been due to the presence of a low mass AGN.

Recently, \citet{Secrest2020AGalaxies} undertook a search for variable AGN in the mid--IR. This was motivated by the sensitivity of mid--IR studies to low luminosity AGN which are frequently optically obscured and Compton--thick \citep{Ricci2016NuSTAR6286, Annuar2017NuSTAR} and frequently unobservable in the soft X-rays \citep{Polimera2018MorphologiesGalaxies}, along with the low contamination rates of supernovae due to weak mid--IR emission \citep{2016PhDT.......177S}. To produce mid--IR light curves, \citet{Secrest2020AGalaxies} used multi--epoch photometry from the Wide--field Infrared Survey Explorer \citep[\textit{WISE}]{Wright2010ThePerformance}. \textit{WISE} mapped the sky in the W1 (3.4$\mu$m) and W2 (4.6$\mu$m) bands with 6 month cadence over an 8 year baseline between the initial observations in 2010, the Near--Earth Object Wide--field Infrared Survey Explorer mission from 2010--2011 \citep[\textit{NEOWISE};][]{Mainzer2011PreliminaryScience} and the reactivation mission beginning in 2013 \citep[\textit{NEOWISE}--R;][]{Mainzer2014InitialMission}. \citet{Secrest2020AGalaxies} found a sample of 2,199 dwarf galaxies of stellar mass $M_* < 2\times 10^9 M_{\odot}\ $ and redshift $0.02<z<0.03$ from the NASA--Sloan Atlas which had corresponding sources in the All\textit{WISE} catalog. Amongst this sample, only 2 (0.09\%) showed significant variability in light curves produced by the All\textit{WISE} Multiepoch Photometry Table and the \textit{NEOWISE}--R Single Exposure Source Table. 

Variability--based search strategies have been particularly good at finding AGN candidates in dwarf galaxies which are optically obscured or unidentifiable by their spectroscopic signatures. For example, only 25\% of the optically variable AGN candidates in galaxies of mass $M_*<10^{10}M_\odot$ found in PTF were classified as AGN or Composite galaxies based on their narrow emission lines \citep{Baldassare2020AFactory}. This is likely due to a combination of star formation dilution of AGN emission lines and the hardening of the accretion disk SED around lower mass BHs which extends the partially ionized zone and reduces the emission line ratios normally used for AGN classification \citep{Cann2019TheRegime}. Variability--based strategies therefore have an important place for finding AGN which would otherwise be missed due to biases in other selection techniques. 

Previous searches for active IMBH candidates in dwarf galaxies with spectroscopic, radio, X-ray and mid--IR observations have not provided a complete, unbiased census of the IMBH occupation fraction because different strategies are hindered by high star formation rates, obscuration and low accretion luminosities. Therefore, despite the discoveries of these large samples of dwarf AGN candidates, there are very few confirmed IMBHs with well--sampled SEDs and measured virial black hole masses which we can use to occupy the sparsely populated low mass ends of a number of black hole--galaxy scaling relations. This motivates the development of effective search strategies for identifying substantially larger samples of IMBH candidates. This will provide more opportunities for careful confirmation and multi--wavelength characterization of the best candidates, especially those with broad Balmer lines.

In this paper we present a comprehensive search for AGN--like variability from a large sample of dwarf galaxies in the optical and mid--IR in order to build upon the previous successes of \citet{Baldassare2018,Baldassare2020AFactory} and \citet{Secrest2020AGalaxies} with SDSS, PTF and \textit{WISE}. For our optical search we have used observations from the Zwicky Transient Facility \citep[ZTF;][]{Bellm2019,Graham2019,Dekany2020TheSystem}, a new and ongoing optical survey which began in March 2018 and achieves single epoch limiting magnitudes of $\sim21$ in g-, r- and i-band over a survey footprint of 23,675 deg$^2$. The Northern Sky Survey of ZTF Phase I (Mar 2018--Sep 2020) had an average cadence of 3 days and this was supplemented by higher cadence sub--surveys with hourly to 1 day cadences \citep{Bellm2019TheScheduler}. The ongoing ZTF Phase II (Oct 2020--present) Northern Sky Survey has a 2 day cadence. By comparison, PTF had a footprint of $\sim8000$ deg$^2$ and a $\sim5$ day cadence. To expand upon the mid--IR variability search of \citet{Secrest2020AGalaxies}, we have made use of new forward modeled photometry catalogs of time--resolved \textit{WISE} coadds to produce more sensitive photometry of a larger dwarf galaxy sample. We do this work in preparation for the upcoming Legacy Survey of Space and Time (LSST) at Vera C. Rubin Observatory \citep{Ivezic2019LSST:Products} which will provide a more complete census of variable IMBHs over the next decade due to its expected single visit limiting magnitude of g $\sim$25 at a 3 day cadence spanning $\sim$10 years. 

In Section 2 we describe our dwarf galaxy sample selection process. In Section 3 we describe our procedure for ZTF forced photometry of the dwarf galaxy sample. In Section 4 we describe our selection strategy for finding the optically variable AGN. In Section 5 we present our sample of IMBH candidates and supernovae from ZTF and describe their multiwavelength and spectroscopic properties. In Section 6 we describe our selection of variable AGN in \textit{WISE} based on forward modeled photometry of time resolved coadds and in Section 7 we discuss the multi--wavelength and spectroscopic properties of the \textit{WISE}--selected IMBH candidates. In Section 8 we discuss the merits of the two selection strategies and the properties of the IMBH candidates in further depth.

\section{Dwarf galaxy sample selection}
We selected a sample of dwarf galaxies using the NASA--Sloan Atlas (NSA) version \texttt{v$_{1\textunderscore0\textunderscore1}$} \footnote{https://www.sdss.org/dr16/manga/manga-target-selection/nsa}. The NSA produced stellar mass estimates for galaxies by fitting 5 templates derived from stellar population synthesis models \citep{Bruzual2003Stellar2003, Blanton2007Near-Infrared} to SDSS images of galaxies after subtracting the sky background \citep{Blanton2011ImprovedImages}. We used this catalog of stellar masses to allow for more direct comparison to other AGN variability searches which classify dwarf galaxies using stellar masses from this catalog \citep[e.g.][]{Reines2013DWARFHOLES, Baldassare2018,Baldassare2020AFactory,Secrest2020AGalaxies}.

When compiling our list of dwarf galaxies from the NSA we required redshifts of $0.02<z<0.35$ and elliptical Petrosian masses of $M_*<3\times h^{-2} 10^9 M_{\odot}$. We selected an $h$ value of 0.73 for consistency with \citet{Reines2013DWARFHOLES} and \citet{Secrest2020AGalaxies} such that our mass cutoff corresponds to $M_*<10^{9.75} M_{\odot}$. After finding some high redshift quasars listed in the catalog with erroneous redshifts and underestimated stellar masses, we required the NSA redshift to be derived from an SDSS spectrum (described by the \texttt{ZSRC} column in the NSA table). This resulted in a final sample of 81,462 dwarf galaxies. 

For the ZTF photometry, we also required that the objects overlap with $>100$ high quality g and r band ZTF images over a $>1$ year baseline. Due to computational limitations, we selected a random subset of the parent light curve sample consisting of 25,714 dwarf galaxies for our ZTF variability search. For our \textit{WISE} variability search we required there to be $>5$ epochs over a $>3$ year baseline and this resulted in a final sample of 79,879 objects for the mid--IR search. 
\begin{figure*}
\gridline{\fig{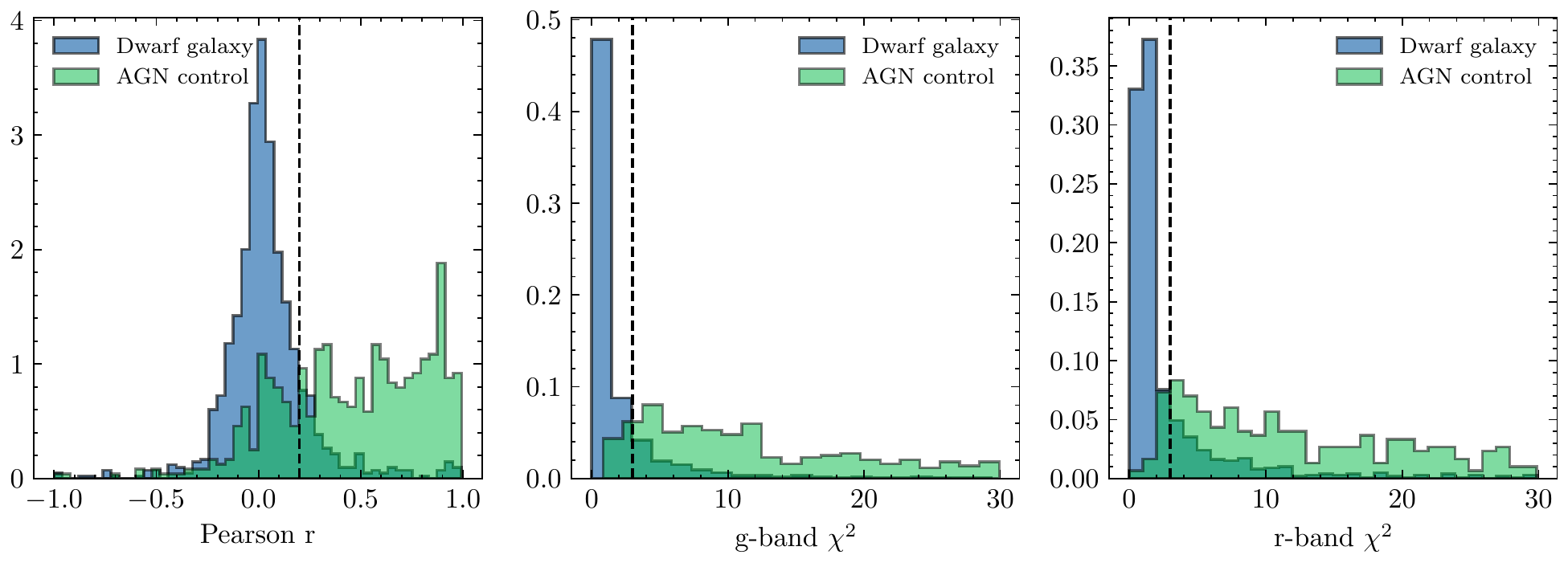}{0.95\textwidth}{\textbf{5 day binning}}}
\gridline{\fig{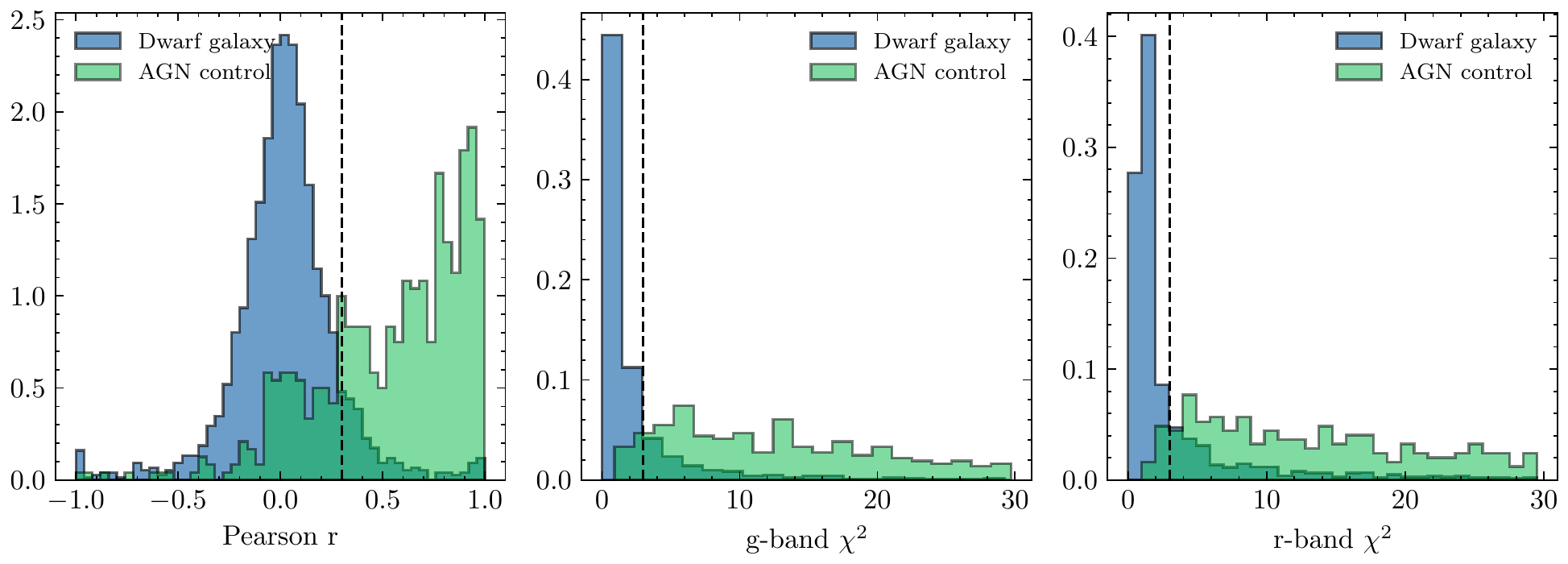}{0.95\textwidth}{\textbf{10 day binning}}}
\gridline{\fig{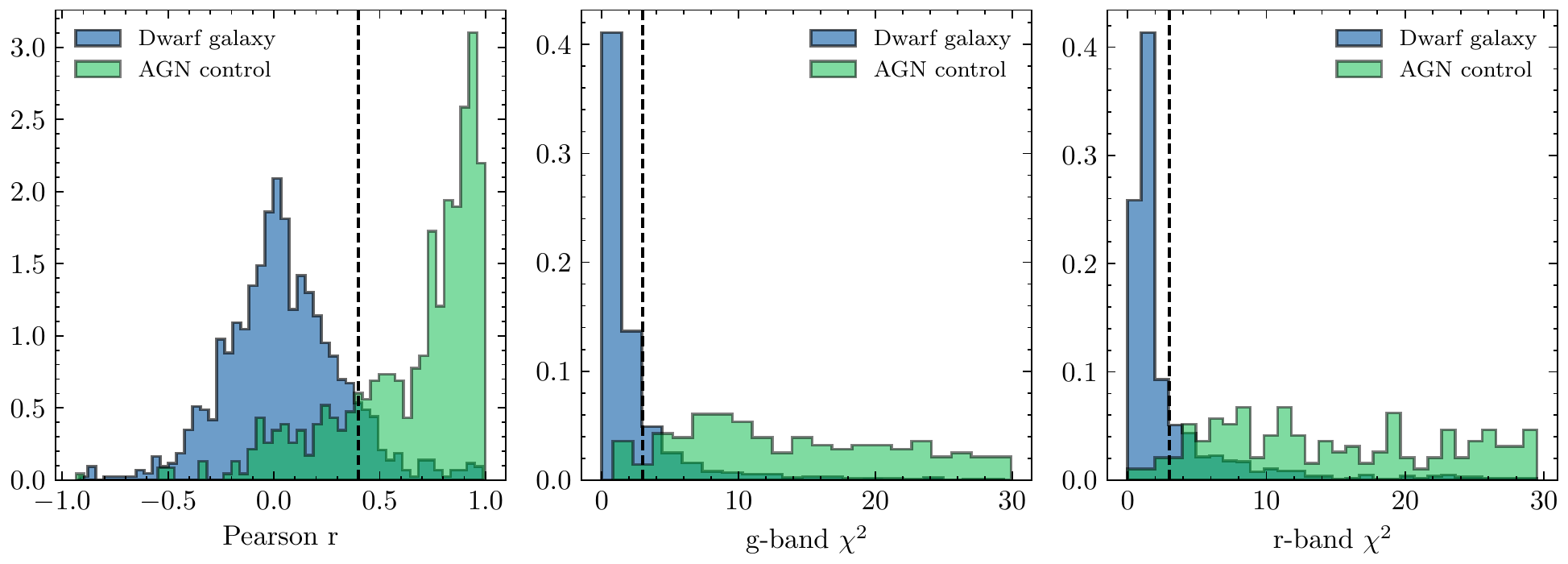}{0.95\textwidth}{\textbf{20 day binning}}}
\caption{Pearson $r$ correlation coefficient and $\chi^2/N$ in g- and r-band calculated from ZTF light curves for 3 different binning timescales. We show the entire dwarf galaxy population with ZTF photometry in blue and the AGN control sample (from host galaxies of stellar mass $M_*>3\times10^{9.75}M_\odot$) in green. The cutoffs used for AGN candidate selection are shown in black dotted lines for each statistic. We required the 3 statistics to satisfy the cutoffs for all 3 binning timescales in order for a candidate to be selected.}
\label{fig:varstat}
\end{figure*}

We produced a control sample of optically variable AGN with host galaxies of mass $M_*>10^{9.75} M_{\odot}\ $ from a parent sample of 5,493 variable ZTF AGN found by \citet{Ward2021AGNsFacility}. This AGN sample was obtained from the ZTF alert stream \citep{Patterson2019} using the AMPEL alert broker \texttt{AMPEL} \citep[Alert Management, Photometry and Evaluation of Lightcurves;][]{Nordin2019} to crossmatch ZTF transients to AGN catalogs using \texttt{catsHTM} \citep{Soumagnac2018} and check for AGN--like variability with the \citet{Butler2011} quasar modeling routine. We matched the AGN from this sample to the NASA--Sloan Atlas with a 5" radius and found 1,053 objects with measured galaxy masses. Both the control sample and the dwarf galaxy sample were processed by the following photometry pipeline.

\section{ZTF photometry of dwarf galaxies}
Difference imaging to detect variability requires the production of reference images, which are co--added stacks of exposure images to support image differencing downstream. To produce new ZTF forced photometry with custom, deeper references than the main ZTF alert pipeline we implemented the ZUDS photometry pipeline\footnote{https://github.com/zuds-survey/zuds-pipeline} on our dwarf galaxy and AGN control samples.
We generated reference images for each field, CCD, quadrant and filter from ZTF single epoch science images by selecting up to 60 images as close as possible in time which met the following criteria established by \citet{Masci2019} for the main ZTF alert pipeline:
\begin{enumerate}
\item Seeing within the range $2."0 \leq \text{FWHM} \leq 5."0$, with priority given to images with lower seeing values. 
\item Quality status $=1$ 
\item Magnitude zeropoints given by $25.3 \leq \text{MAGZP(g)} \leq 26.5$ or $25.3 \leq \text{MAGZP(r)} \leq 26.5$ for filters g and r respectively.
\item Color coefficients given by $-0.20 \leq \text{CLRCOEFF(g)} \leq 0.15$ or $-0.05 \leq \text{CLRCOEFF(r)} \leq 0.22$ for filters g and r respectively.
\item Limiting magnitudes given by $\text{MAGLIM(g)} \geq 19.0$ or $\text{MAGLIM(r)} \geq 19.0$ for filters g and r respectively.
\item Global pixel median $\leq 1900$ DN or $\leq 1600$ for filters g and r respectively.
\item Global robust pixel RMS $\leq 100$ DN.
\item Acquired after camera reinstallation on February 5, 2018 UT. 
\end{enumerate}

When there were multiple field, CCD and quadrants containing the source for a particular filter, we selected the reference image containing the largest number of high quality ZTF science images in the coadd to be the reference image. We produced 1000"x1000" cutouts of each single epoch science image and reference image, produced subtractions, then undertook aperture photometry with a 3".0 radius. We then applied the aperture to PSF correction factors produced by the main ZTF image pipeline to produce PSF photometry light curves. We measured the baseline flux of each object in the reference image and added this to the fluxes measured from the image subtractions.
\begin{figure}
\gridline{\fig{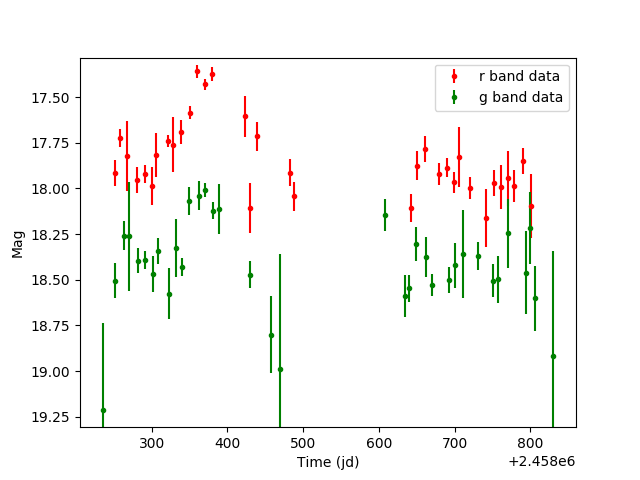}{0.45\textwidth}{NSA189758}}
\gridline{ \fig{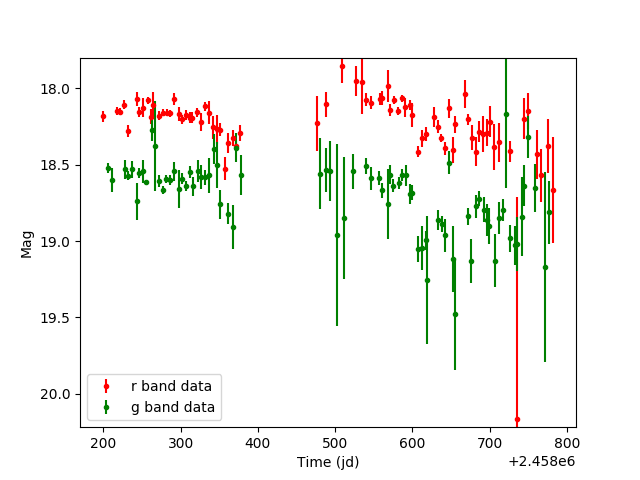}{0.45\textwidth}{NSA212423}}
\gridline{\fig{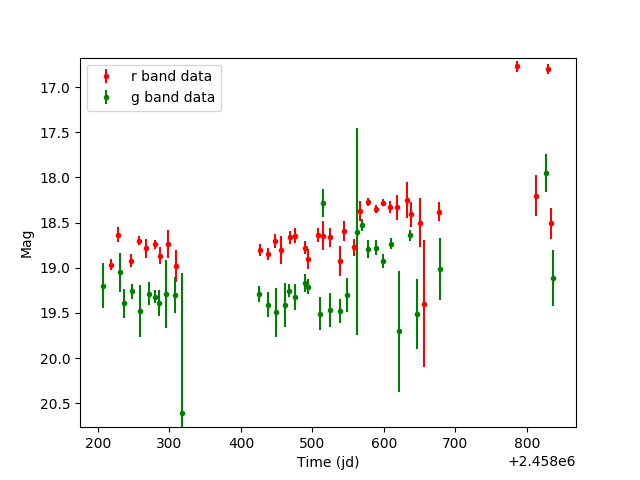}{0.45\textwidth}{NSA158589}}
\caption{Three examples of ZTF light curves which passed the variability criteria and were classified as variable AGN candidates.}
\label{fig:lcs}
\end{figure}

In order to improve our sensitivity to low S/N variability from faintly varying AGN and prepare the data for calculation of the Pearson correlation coefficient between g- and r-band photometry, we binned the data in temporal bins. We first applied zeropoint corrections, then undertook error--weighted binning in flux space using bins of 5, 10 and 20 day increments. Therefore, each galaxy had 3 light curves with different bin sizes for the calculation of variability statistics. This binning procedure may have reduced our sensitivity to objects with optical variability only on timescales $<$20 days and therefore may have introduced biases against particularly low mass AGN. However, we determined that binning was necessary to both allow for the calculation of the Pearson r correlation coefficient whilst confidently identifying light curves with correlated variability over a range of timescales, as expected from AGN power spectra \citep{Kelly2009AreFluctuations,MacLeod2011QUASARVARIABILITY,Burke2021ADisks} With this procedure we generated light curves of 25,714 dwarf galaxies and the 1,053 AGN from the high mass galaxy control sample.

\section{Selection of variable IMBH candidates}
In order to detect real variability amongst the sample of dwarf galaxy light curves we trialed a number of variability statistics including the Pearson correlation coefficient (as used by \citet{Secrest2020AGalaxies}), the goodness of fit of the light curves to the quasar structure function (using \texttt{qsofit} \citep{Butler2011}, as used in \citet{Baldassare2018, Baldassare2020AFactory,Ward2021AGNsFacility}), the excess variance \citep{Sanchez2017Near-infraredField,Martinez-Palomera2020IntroducingSurvey} and the $\chi^2/N$ in g and r bands. We found that a combination of the Pearson correlation coefficient and $\chi^2$ produced the best separation between the AGN control sample and the majority of the non--variable dwarf galaxy population.

The Pearson correlation coefficient $r$ between the binned g and r band fluxes was calculated as:
\begin{equation}
    r=\frac{C_{f_g,f_r}}{\sigma_{f_g}\sigma_{f_r}}
\end{equation}
where $C_{f_g,f_r}$ is the covariance between g and r bands given by
\begin{equation}
    C_{f_g,f_r}=\frac{1}{N-1}\sum^{N}_i(f_{g,i}-\langle f_g \rangle)\times (f_{r,i}-\langle f_r \rangle ) 
\end{equation}
and $\sigma_{f_g}$ and $\sigma_{f_r}$ are the g and r band variability amplitudes given by:
\begin{equation}
    \sigma^2_f=\frac{1}{N-1}\sum^{n}_i(f_i-\langle f \rangle )^2
\end{equation}
and the expectation value $\langle f \rangle$ was given by the median flux. 

We also calculated the $\chi^2/N$ of the light curves in g and r bands:

\begin{equation}
    \chi^2/N=\frac{1}{N}\sum^{N}_i\frac{(f_{i}-\langle f \rangle)^2}{\sigma_{i}^2}
\end{equation}
The distribution of the Pearson correlation coefficient $r$ and the $\chi^2/N$ values for g and r bands for the dwarf galaxy and control AGN light curves with 5, 10 and 20 day bin sizes is shown in Figure \ref{fig:varstat}. We applied cutoffs of $r>0.2$, $r>0.3$, and $r>0.4$ for 5, 10 and 20 day bin sizes respectively, and $\chi^2/N>3$ in both filters for all bin sizes, to classify dwarf galaxies as variable. Each dwarf galaxy was required to meet this cutoff in all 3 binning timescales to ensure that correlations and variance was not produced as artifacts of bin phase and size. 
\begin{figure*}[h]
\gridline{\fig{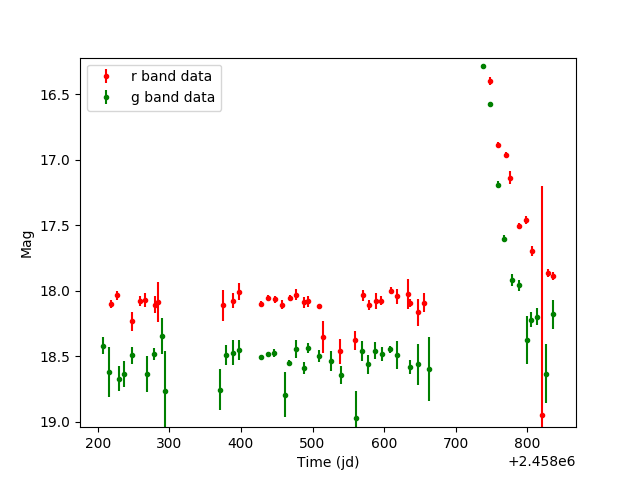}{0.32\textwidth}{NSA164967}
 \fig{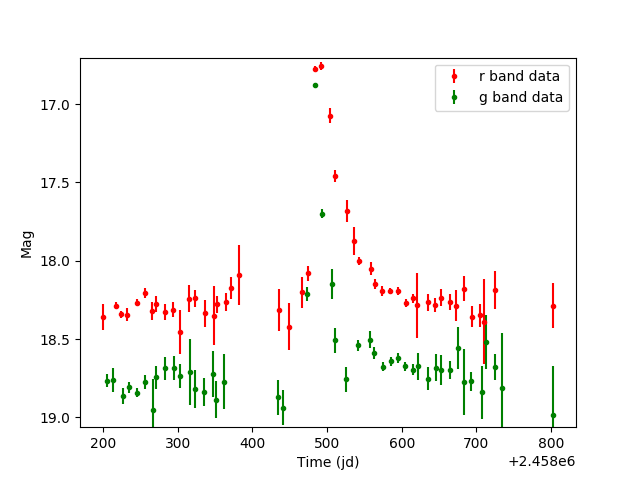}{0.32\textwidth}{NSA136891} \fig{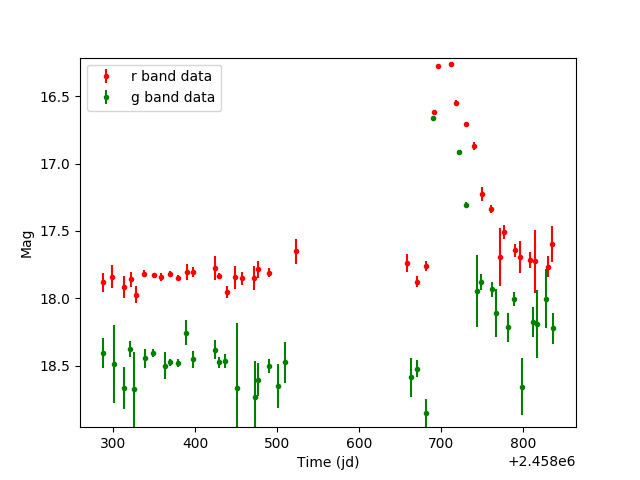}{0.32\textwidth}{NSA33420}}
\caption{Three examples of ZTF light curves which passed the variability criteria and were classified as supernova candidates.}
\label{fig:lcsSN}
\end{figure*}
After applying these cutoffs, we found 130 dwarf galaxies with statistically significant variability. 

We then manually inspected the light curves and difference images to remove light curves with high variance due to poor quality photometry and to determine which light curves contained single flares with reddening consistent with supernovae. 36 supernovae--like flares were found, corresponding to a nuclear SN rate of $0.14\pm0.02$\% ($0.05\pm0.01$\% year$^{-1}$) for dwarf galaxies in ZTF Phase I.

This process produced a final sample of 44 dwarf galaxies ($0.17\pm0.03$\%) with variability consistent with AGN activity. These constitute our set of optically variable IMBH candidates. The properties of this sample are summarized in Table \ref{table:ztfcands}. Examples of ZTF light curves of 3 dwarf galaxies with AGN--like variability are shown in Figure \ref{fig:lcs} and examples of 3 supernova light curves are shown in Figure \ref{fig:lcsSN}.

\begin{deluxetable*}{rcccccccc}
\tabletypesize{\scriptsize}
\tablecolumns{9}
\tablewidth{0pt}
\tablecaption{Properties of the ZTF-selected IMBH candidates \label{table:ztfcands}}
\tablehead{
\colhead{NSA ID} & \colhead{RA (hms) } & \colhead{Dec (dms)} & \colhead{z} &\colhead{log$_{10}$M$_*$($M_\odot$)}&\colhead{log$_{10}$M$_{\text{BH}}$($M_\odot$)[M$_*$]}& \colhead{BPT class} &\colhead{BLR?}&\colhead{Virial log$_{10}M_{\text{BH}}(M_\odot$)} }
\startdata
8479&12:31:07.418&00:27:47.447&$0.0214$&$9.51$&$6.95$&SF&$\times$&-\\
19864&12:36:43.723&-3:01:15.176&$0.0082$&$8.12$&$5.23$&SF&$\times$&-\\
20245&12:42:32.662&-1:21:02.583&$0.0037$&$8.75$&$6.01$&SF&$\times$&-\\
23518&16:21:41.608&01:02:53.989&$0.0324$&$9.57$&$7.03$&SF&$\times$&-\\
30890&23:18:28.463&00:20:47.845&$0.0341$&$9.66$&$7.14$&Composite&$\times$&-\\
32653&00:09:26.408&00:19:32.182&$0.1138$&$9.74$&$7.24$&Seyfert&\checkmark&7.048\\
35747&01:10:59.314&00:26:01.103&$0.0188$&$9.7$&$7.19$&Seyfert&\checkmark&7.561\\
37197&01:44:32.794&00:57:15.492&$0.0716$&$9.59$&$7.05$&SF&$\times$&-\\
42027&00:43:38.702&14:38:29.188&$0.0766$&$9.74$&$7.24$&SF&$\times$&-\\
43101&01:06:29.847&14:58:17.054&$0.0393$&$9.54$&$6.99$&SF&$\times$&-\\
44934&02:11:58.218&14:16:09.393&$0.0264$&$9.45$&$6.88$&SF&$\times$&-\\
46229&07:47:06.104&37:36:55.296&$0.0364$&$9.54$&$6.99$&SF&$\times$&-\\
49405&08:29:12.677&50:06:052.41&$0.0432$&$9.54$&$6.99$&SF&\checkmark&6.363\\
52237&09:16:08.642&55:59:22.952&$0.0721$&$9.67$&$7.15$&SF&$\times$&-\\
78310&09:05:10.183&50:42:10.045&$0.0369$&$9.74$&$7.24$&SF&$\times$&-\\
83637&10:19:48.118&03:31:18.466&$0.0324$&$9.05$&$6.38$&SF&$\times$&-\\
95377&15:23:50.527&56:48:15.474&$0.0511$&$9.41$&$6.83$&SF&$\times$&-\\
97216&16:05:40.674&50:45:16.842&$0.0129$&$8.2$&$5.33$&SF&$\times$&-\\
115553&02:54:57.457&00:25:55.471&$0.1077$&$9.47$&$6.9$&SF&$\times$&-\\
117445&22:16:19.958&-7:20:55.229&$0.0591$&$9.15$&$6.51$&SF&$\times$&-\\
127775&00:17:18.375&15:49:06.185&$0.0603$&$9.36$&$6.77$&SF&$\times$&-\\
132743&09:49:53.016&60:53:25.584&$0.0828$&$9.57$&$7.03$&SF&$\times$&-\\
133634&11:00:44.374&60:20:05.121&$0.0171$&$8.46$&$5.65$&SF&$\times$&-\\
138748&14:55:45.198&57:20:49.518&$0.0795$&$9.56$&$7.01$&SF&$\times$&-\\
142229&15:57:46.352&47:17:47.018&$0.0316$&$9.35$&$6.75$&SF&$\times$&-\\
142571&16:20:58.965&45:10:54.736&$0.0542$&$9.51$&$6.95$&Composite&$\times$&-\\
158589&11:30:27.424&52:18:13.171&$0.0483$&$9.37$&$6.78$&SF&$\times$&-\\
159003&11:41:52.296&52:38:24.914&$0.0191$&$8.91$&$6.21$&SF&$\times$&-\\
164884&09:28:01.292&49:18:17.303&$0.1143$&$9.71$&$7.2$&SF&\checkmark&6.961\\
168943&13:45:010.34&-3:01:20.764&$0.0469$&$9.42$&$6.84$&SF&$\times$&-\\
181600&13:06:15.181&58:57:0033.1&$0.0292$&$9.64$&$7.11$&Seyfert&\checkmark&6.732\\
183841&11:09:18.046&49:47:53.629&$0.0466$&$9.3$&$6.69$&SF&$\times$&-\\
188282&17:15:34.141&31:31:45.348&$0.0229$&$9.61$&$7.08$&SF&$\times$&-\\
189758&21:05:04.972&00:01:010.03&$0.0312$&$9.46$&$6.89$&Composite&$\times$&-\\
197553&10:57:14.371&54:10:56.999&$0.072$&$9.34$&$6.74$&SF&$\times$&-\\
198091&11:32:00.219&53:42:50.763&$0.0268$&$9.62$&$7.09$&SF&$\times$&-\\
202748&21:24:056.67&00:38:45.601&$0.048$&$9.19$&$6.56$&SF&$\times$&-\\
212423&16:18:03.954&35:08:08.626&$0.033$&$9.59$&$7.05$&SF&$\times$&-\\
272369&11:38:55.181&44:19:031.64&$0.0321$&$9.45$&$6.88$&Composite&$\times$&-\\
277358&14:52:32.685&36:24:47.857&$0.0542$&$9.52$&$6.96$&SF&$\times$&-\\
298494&12:19:51.136&47:50:37.875&$0.0384$&$9.71$&$7.2$&Seyfert&$\times$&-\\
301868&13:25:45.394&46:07:01.166&$0.0365$&$9.25$&$6.63$&SF&$\times$&-\\
451469&14:14:05.019&26:33:36.804&$0.0358$&$9.66$&$7.14$&SF&\checkmark&6.303\\
545880&13:37:02.948&18:10:13.979&$0.0265$&$8.82$&$6.1$&SF&$\times$&-\\
\enddata
\vspace{0.1cm}
\tablecomments{Properties of the 44 AGN candidates with significant and correlated g and r band variability found in ZTF. The IDs, positions, redshifts and host galaxy stellar masses are those from the NSA catalog version \texttt{v$_{1\textunderscore0\textunderscore1}$}. In the first log$_{10}$M$_{\text{BH}}$ column we show the estimated black hole mass based on the $M_*-M_{\text{BH}}$ from \citet{Schutte2019TheNuclei} which has a scatter of 0.68 dex. The presence of Balmer broad lines is indicated in the BLR column. Virial masses were calculated for broad line AGN by \citet{Ho2015ASPECTRA} based on the width of the H$\alpha$ broad lines.}
\end{deluxetable*}

\begin{deluxetable*}{rccccccc}
\tabletypesize{\scriptsize}
\tablecolumns{8}
\tablewidth{0pt}
\tablecaption{Properties of the ZTF--selected supernovae in dwarf galaxies \label{table:SNcands}}
\tablehead{
\colhead{NSA ID} & \colhead{RA (hms) } & \colhead{Dec (dms)} & \colhead{z} &\colhead{log$_{10}$M$_*$($M_\odot$)}& \colhead{BPT class} &\colhead{BLR?}&\colhead{Classification}}
\startdata
6971&12:00:49.206&-1:15:01.297&$0.021$&$8.55$&SF&$\times$&None\\
8600&12:35:43.642&00:13:25.722&$0.0231$&$9.43$&Seyfert&$\times$&None\\
14878&14:48:27.429&00:12:04.871&$0.0269$&$8.94$&SF&$\times$&None\\
33420&00:28:51.572&-1:04:22.318&$0.0333$&$9.25$&SF&$\times$&SNIa\\
38750&02:33:46.936&-1:01:28.378&$0.0489$&$9.51$&SF&$\times$&None\\
41092&03:35:26.631&00:38:11.445&$0.0232$&$8.23$&SF&$\times$&None\\
47343&08:00:26.505&42:49:16.255&$0.0408$&$9.58$&SF&$\times$&SNIa\\
60574&10:26:03.501&64:50:10.285&$0.0368$&$9.47$&SF&$\times$&SNIa\\
60660&10:40:58.604&65:30:54.366&$0.0108$&$8.25$&SF&$\times$&SNIc\\
70478&13:02:24.185&03:22:032.91&$0.0234$&$9.25$&SF&$\times$&None\\
77819&08:35:16.365&48:19:01.178&$0.0429$&$9.55$&SF&$\times$&SNIc\\
81368&08:55:49.031&03:38:32.333&$0.0274$&$9.59$&SF&$\times$&SNIIn\\
90779&11:11:03.618&64:14:46.086&$0.0352$&$9.46$&SF&$\times$&SNII\\
121437&21:51:07.575&12:25:36.661&$0.0285$&$9.46$&SF&$\times$&SNIc-BL\\
126810&23:58:48.168&14:44:29.893&$0.0274$&$8.95$&SF&$\times$&SNII\\
136564&13:21:28.156&62:52:029.91&$0.0413$&$9.66$&SF&$\times$&SNIa\\
136891&13:29:31.023&61:26:000.53&$0.0313$&$9.25$&SF&$\times$&None\\
137293&14:02:21.316&59:25:08.501&$0.0321$&$9.61$&SF&$\times$&None\\
137916&14:17:15.746&59:16:39.876&$0.0365$&$9.24$&SF&$\times$&SNIa\\
141203&03:26:45.983&01:11:01.017&$0.0485$&$9.13$&SF&$\times$&SNIa\\
146295&11:19:04.838&03:42:53.379&$0.0273$&$8.74$&SF&$\times$&None\\
164967&09:35:11.377&48:43:08.652&$0.0244$&$8.9$&SF&$\times$&SNIa\\
183178&10:43:31.995&48:26:37.354&$0.0413$&$9.44$&SF&$\times$&SNIa\\
191873&09:36:08.603&06:15:25.511&$0.008$&$8.43$&SF&$\times$&SN\\
208509&14:20:52.371&50:31:28.532&$0.0446$&$9.52$&SF&$\times$&SNII\\
211965&16:12:50.281&36:01:18.328&$0.0314$&$9.09$&SF&$\times$&None\\
214163&02:36:15.641&-1:11:59.948&$0.0471$&$9.2$&SF&$\times$&SNIa\\
216212&00:13:45.056&00:51:05.651&$0.0322$&$8.55$&SF&$\times$&None\\
220206&14:49:48.132&54:48:46.962&$0.0448$&$9.47$&SF&$\times$&SNIa\\
224290&16:57:15.706&31:57:50.583&$0.0377$&$9.3$&SF&$\times$&SNII\\
229236&09:04:08.208&39:40:35.155&$0.028$&$9.5$&SF&$\times$&SNIa\\
249950&15:08:50.831&40:42:44.478&$0.0305$&$9.38$&SF&$\times$&SNII\\
269844&10:23:20.137&41:23:40.255&$0.0515$&$9.6$&SF&$\times$&SNIa\\
271156&11:02:16.807&44:20:06.661&$0.0246$&$9.64$&Composite&$\times$&SNII\\
289679&10:15:29.897&37:59:02.588&$0.054$&$8.93$&SF&$\times$&SNIa\\
298529&12:22:36.827&47:58:08.744&$0.0499$&$9.5$&SF&$\times$&SNIa\\
\enddata
\vspace{0.1cm}
\tablecomments{Properties of the 36 supernova candidates with significant and correlated g and r band variability found in ZTF. The IDs, positions, redshifts and host galaxy stellar masses are those from the NSA catalog version \texttt{v$_{1\textunderscore0\textunderscore1}$}. The presence of Balmer broad lines is indicated in the BLR column. The last column shows the spectroscopic classification made published by ZTF on the Transient Name Server.}
\end{deluxetable*}

\begin{figure*}
\gridline{\fig{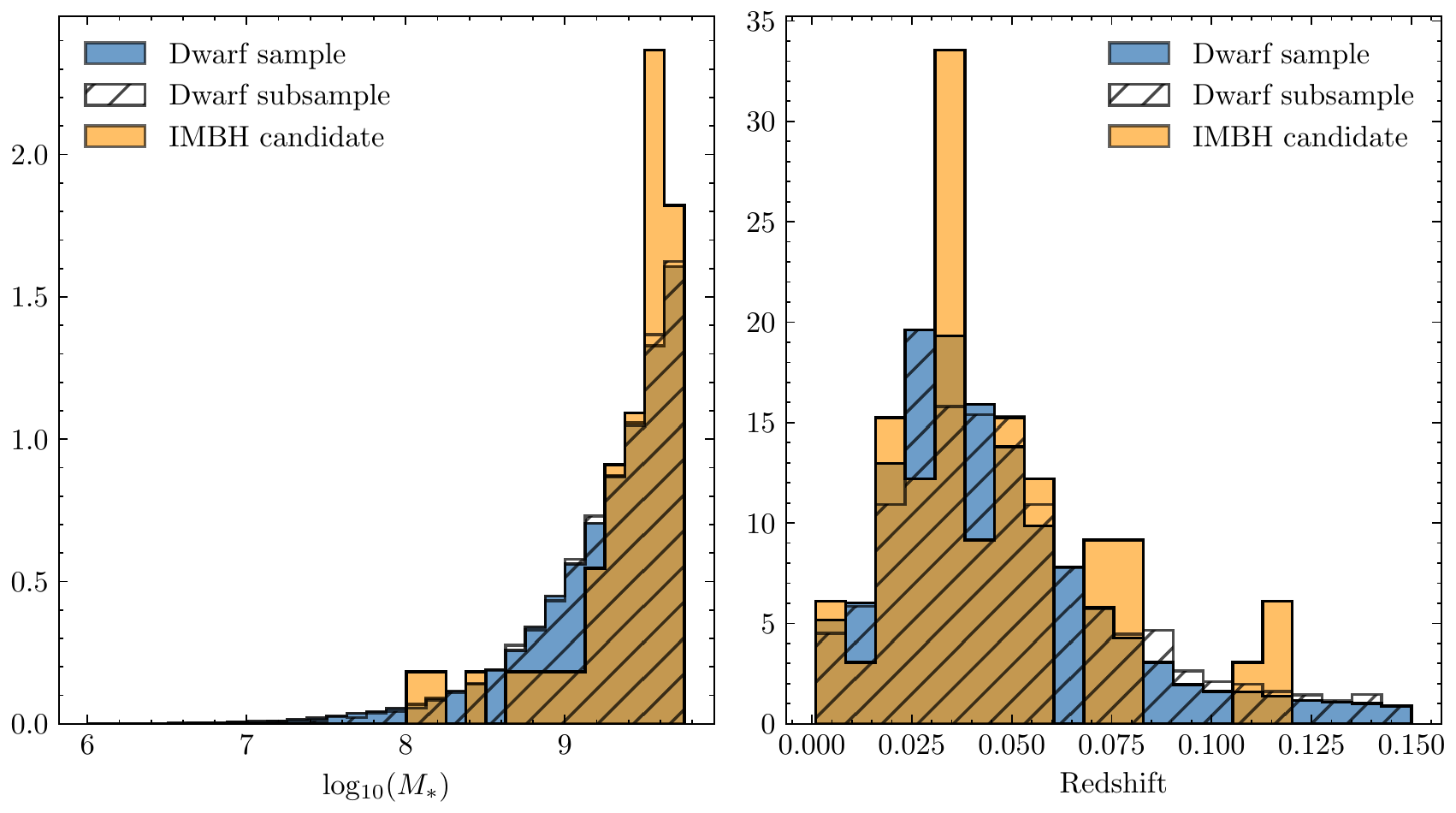}{0.65\textwidth}{AGN candidate sample}}
\gridline{\fig{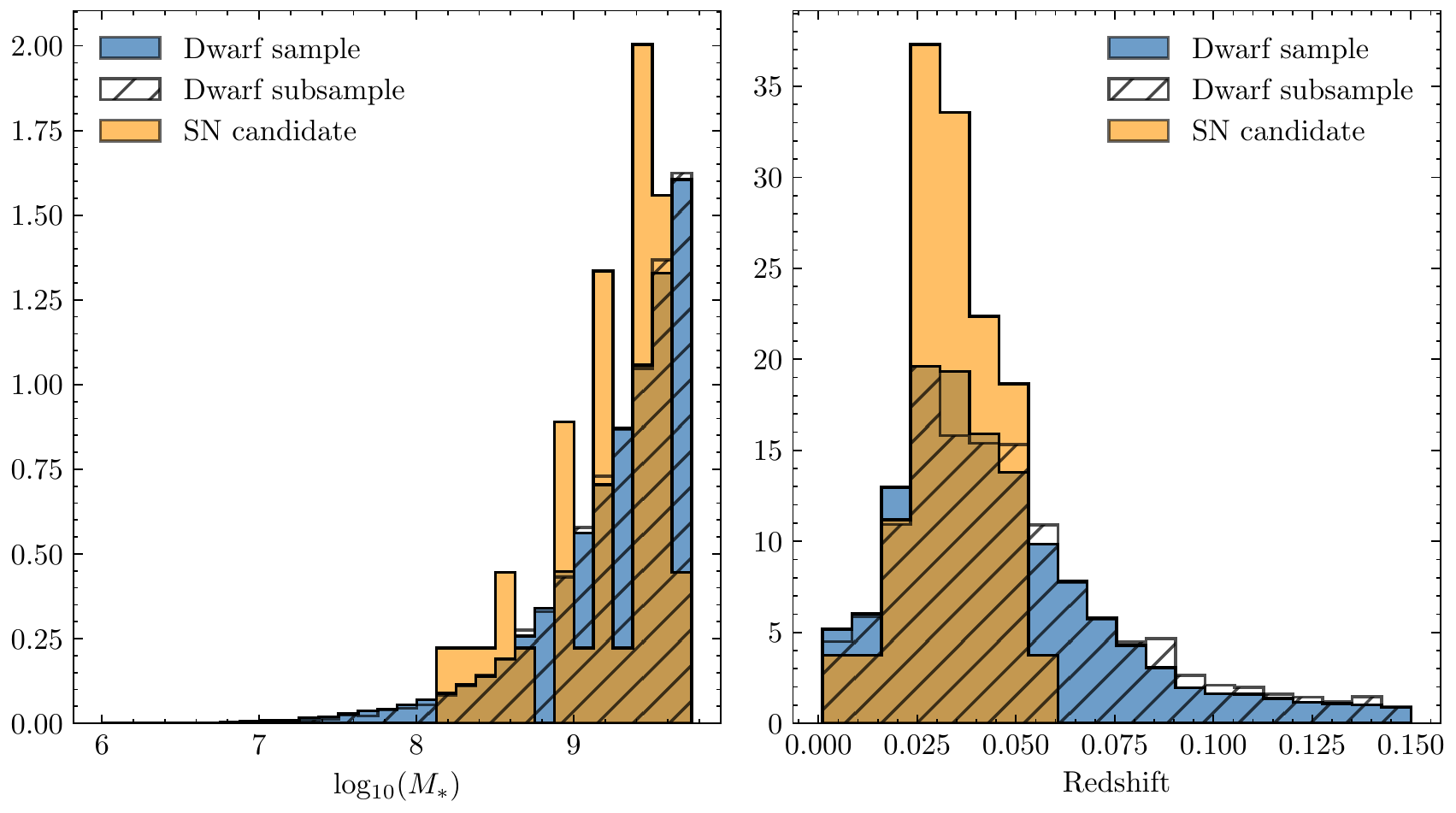}{0.65\textwidth}{Supernova sample}}
\caption{Normalized galaxy mass and redshift distributions of variable IMBH candidates (top) and supernova candidates (bottom) shown in green and the parent dwarf galaxy sample shown in blue. The subset of the parent dwarf galaxy for which ZTF photometry was produced for the optical IMBH search is shown in the hatched histogram. Redshifts and host galaxy mass measurements were derived from the NSA.}
\label{fig:galstat}
\end{figure*}

The distributions of the redshifts and host galaxy stellar masses of the optically variable dwarf galaxies are shown in Figure \ref{fig:galstat}. All but 6 of the IMBH candidates were in galaxies of mass $M_*>10^9M_\odot$. By comparison, a larger fraction of supernovae were observed in low mass galaxies and most were found in a narrow redshift range of $0.02<z<0.055$.

We checked both the supernova and AGN candidate samples for spectroscopic classification on the Transient Name Server\footnote{https://www.wis-tns.org/}. 25 of the 36 supernova candidates had published spectroscopic classifications. One of these was classified as a supernova but was not typed. 17 out of the remaining 24 (71\%) were classified as SNIa, SNIc or SNIc--BL. 7 out of 24 (29\%) were classified as SNII or SNIIn. These classifications are shown in Table \ref{table:SNcands}. We note that the remaining 11 SN without spectroscopic classifications can only be classified as SN candidates, as they may be AGN outbursts \citep{Drake2019ResultsGalaxies}. None of the AGN candidates had spectroscopic classifications published on TNS.

In order to estimate the black hole masses of our IMBH candidates based on their host galaxy stellar mass we used the updated relationship between black hole mass and bulge mass derived by \citet{Schutte2019TheNuclei}. This study found a linear relationship based on a sample of galaxies with carefully measured black hole and bulge masses including 8 dwarf galaxies with stellar masses log$(M_*)<8.5 M_\odot$:
\begin{align}
&\log(M_{\text{BH}}/M_\odot) = \alpha + \beta \log(M_*/10^{11}M_\odot)\\\nonumber&\alpha=8.80\pm0.085; \beta=1.24\pm0.081 
\end{align}
The relation has a scatter of 0.68 dex. The black hole mass estimates for the IMBH candidates are shown in Table \ref{table:ztfcands}.

\section{Spectroscopic and multi--wavelength properties of the ZTF--selected IMBH candidates}
\begin{figure*}
\gridline{\fig{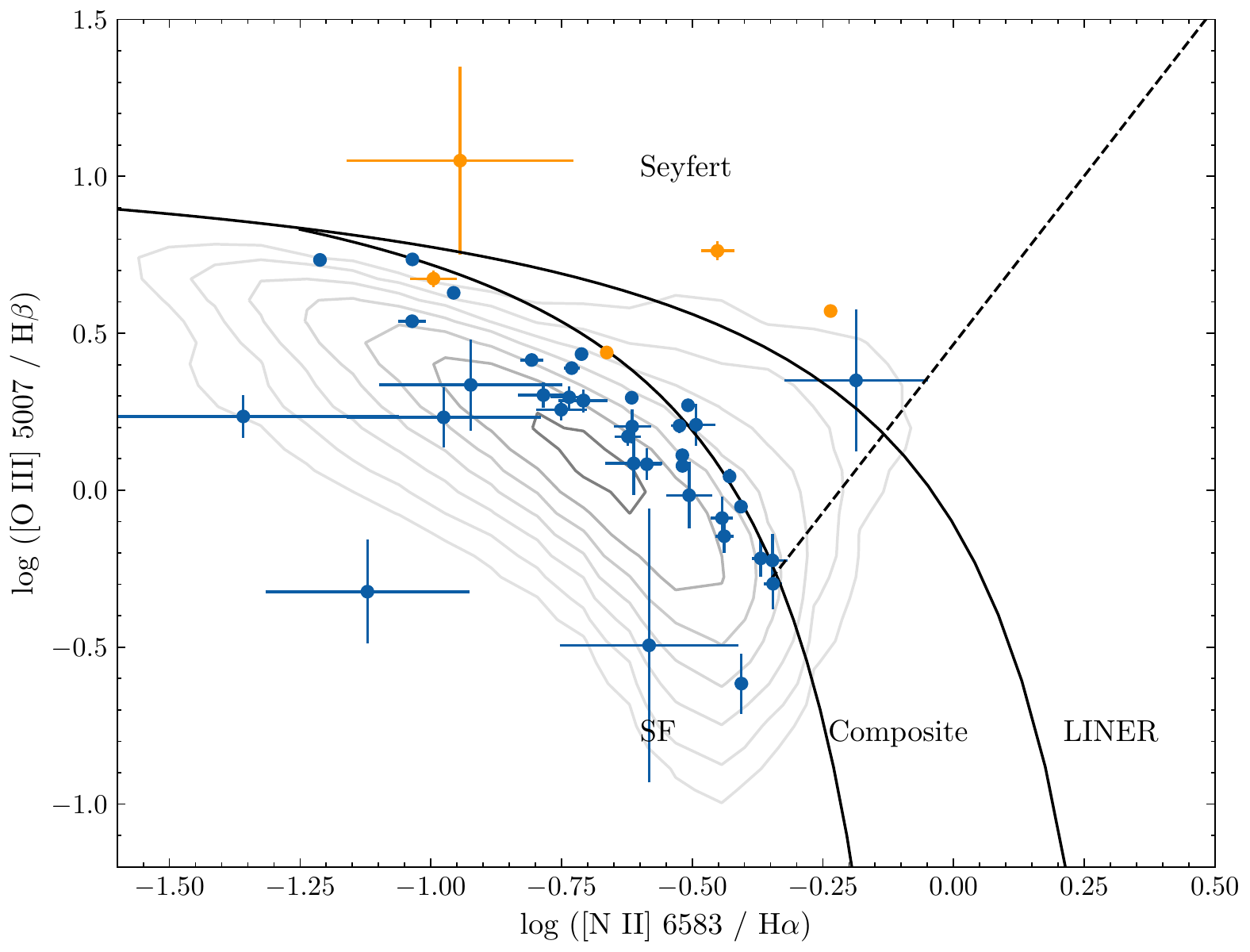}{0.45\textwidth}{AGN candidates}\fig{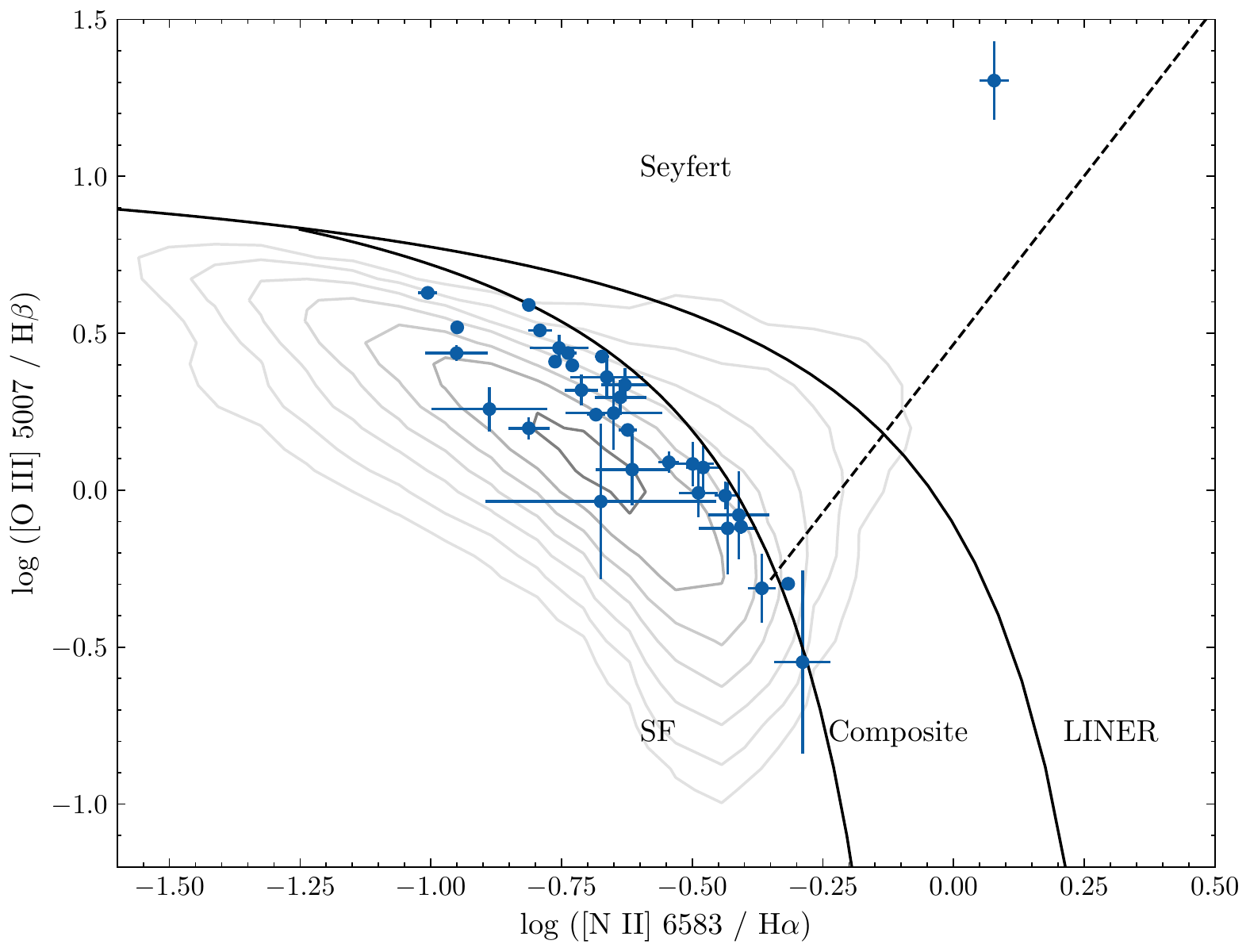}{0.45\textwidth}{Supernova candidates}}
\caption{BPT diagram showing narrow line ratios derived from \texttt{pPXF} fitting of archival SDSS spectra of the ZTF--selected IMBH candidates (left) and supernova candidates (right). Orange points show line ratios for IMBH candidates with broad Balmer lines in archival spectra and blue points show those with narrow emission lines only. Grey contours show the population density with log scaling of the entire parent dwarf galaxy sample for line ratios derived from the NASA--Sloan Atlas. Classification regions are labeled in black text. We note that because \texttt{pPXF} determines narrow emission line strength after accounting for stellar absorption of those same lines, fluxes may appear to be slightly larger in the BPT diagram compared to the NSA--derived emission lines fluxes of the dwarf galaxy population.}
\label{fig:bpt}
\end{figure*}In order to determine the spectroscopic class of the IMBH candidates on the Baldwin, Phillips and Terlevich diagram \citep{Baldwin1981ClassificationObjects,Veilleux1987SpectralGalaxies} we fit the narrow emission line ratios with Penalized Pixel Fitting (pPXF) \citep{Cappellari2003,Cappellari2017}. This method finds the best fit stellar continuum and absorption model using a large sample of high resolution templates of single stellar populations adjusted to match the spectral resolution of the input spectrum. We simultaneously fit the narrow H$\alpha$, H$\beta$, H$\gamma$, H$\delta$,  [S\,{\sc ii}] $\lambda$6717, 6731, [N\,{\sc ii}] $\lambda$6550, 6575, [O\,{\sc i}] $\lambda$6302, 6366 and [O\,{\sc iii}] $\lambda$5007, 4959 narrow emission lines as well as the best fit stellar continuum spectrum during template fitting.

The classification of the IMBH candidates on the [O\,{\sc iii}] $\lambda$5007/H$\beta$ -- [N\,{\sc ii}] $\lambda$6583 /H$\alpha$ BPT diagram is shown in Figure \ref{fig:bpt}. 35 objects (81\%) are classified as star forming while 4 objects (9\%) are in the composite region and 4 (9\%) are classified as Seyferts. For comparison, we also show the density of the original 81,462 dwarf galaxy parent sample on the BPT diagram using emission line ratios quoted in the NSA catalog. From the grey contours it can be seen that the majority of the dwarf galaxy population is star forming but a small population extends out to AGN and LINER regions of the BPT classification scheme. Our typical star forming IMBH candidates lie at higher emission line ratios than this population, in part because our \texttt{pPXF} modeling considers stellar absorption of those lines.

The classification of the supernova host galaxies on the [O\,{\sc iii}] $\lambda$5007/H$\beta$ -- [N\,{\sc ii}] $\lambda$6583 /H$\alpha$ BPT diagram is also shown in Figure \ref{fig:bpt}. 34 host galaxies (94\%) are classified as star forming while 1 object (3\%) was in the composite region and 1 (3\%) in the Seyfert region.

We crossmatched our ZTF--selected IMBH candidates to the active dwarf galaxies from the \citet{Secrest2020AGalaxies} mid--IR variability search, the \citet{Baldassare2018,Baldassare2020AFactory} optical variability searches, the \citet{Molina2021AEvents} [Fe\,X] $\lambda$6374 coronal line emission search, the \citet{Mezcua2020HiddenAGN} IFU spectroscopy search, the \citet{Mezcua2018Intermediate-massSurvey} Chandra X-ray search, the \citet{Latimer2021AGalaxies} mid--IR color selection box search and the \citet{Reines2013DWARFHOLES} optical emission line search. We found that only 7 objects had been detected in previous optical variability searches \citep{Baldassare2018,Baldassare2020AFactory}, where they have different IDs due to the use of NSA version v$_{0\textunderscore1\textunderscore2}$: NSA32653 (SDSS), NSA49405 (PTF), NSA115553 (SDSS), NSA181600 (PTF), NSA202748 (PTF),  NSA451469 (PTF) and NSA545880 (PTF). No other multi--wavelength detections of our candidates from previous IMBH searches were found. 

NSA32653, NSA35747, NSA49405, NSA164884, NSA464884, NSA181600 and NSA451469 were the only objects in our IMBH sample to exhibit broad Balmer lines in archival SDSS spectra. These objects previously had their virial black hole masses estimated using the width of the H$\alpha$ broad lines by \citet{Ho2015ASPECTRA} and \citet{Liu2019ADR7}. These virial masses ranged between $10^{6.3}M_\odot$ and $10^{7.6}M_\odot$ and are shown in Table \ref{table:ztfcands}.

\section{ZTF variability of previously reported PTF--selected IMBH candidates}
After discovering that 5 objects from the variability--selected AGN sample from PTF \citep{Baldassare2020AFactory} were also variable in ZTF according to our selection thresholds, we decided to determine if the remaining AGN candidates in dwarf galaxies from PTF had ZTF variability which was missed by our selection criteria.

We first found that our parent dwarf galaxy sample had 152 objects which overlapped with the variable PTF sample. We then visually inspected the \texttt{zuds-pipeline} light curves we made of the 152 common dwarf galaxies and confirmed that no other dwarf galaxy had apparent variability. We then used the ZTF forced photometry service \citep{Masci2019} to obtain alternative photometry of the sources with the original ZTF reference images. After removing poor quality images by requiring the \texttt{procstatus} flag be $=0$, we measured the baseline flux from  the reference images, applied zeropoints and combined the baseline and single epoch fluxes to produce the light curves of the 152 objects. We then visually inspected the candidates to look for any signs of variability. The alternative pipeline confirmed that only 5 candidates from our overlapping samples showed statistically significant variability in ZTF.

\section{\textit{WISE} single epoch forced photometry}

\begin{figure*}
\gridline{\fig{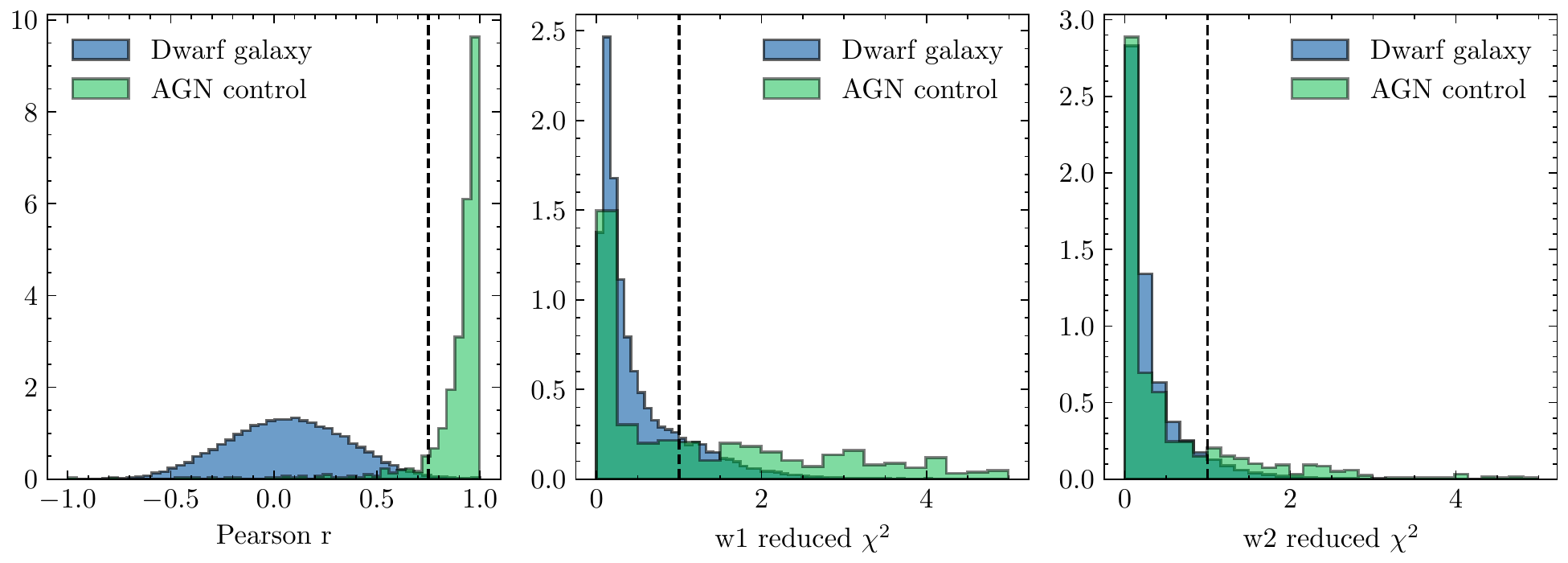}{0.95\textwidth}{}}
\caption{Pearson $r$ correlation coefficient and $\chi^2/N$ in W1 and W2 band calculated from \textit{WISE} forward modeling light curves. We show the entire dwarf galaxy population with \textit{WISE} photometry in blue and the AGN control sample (from host galaxies of stellar mass $M_*>3\times10^{9.75}M_\odot$) in green. The cutoffs used for AGN candidate selection are shown in black dotted lines for each statistic. We required the 3 statistics to satisfy the cutoffs in order for a candidate to be selected.}
\label{fig:varstat_wise}
\end{figure*}

\begin{figure*}[h]
\gridline{\fig{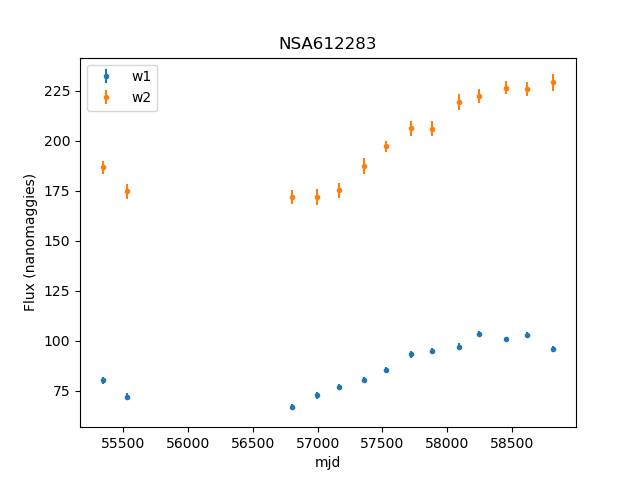}{0.32\textwidth}{NSA612283} 
 \fig{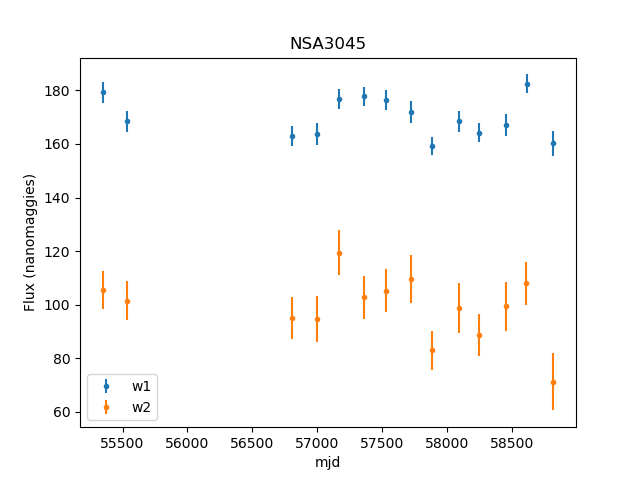}{0.32\textwidth}{NSA3045} \fig{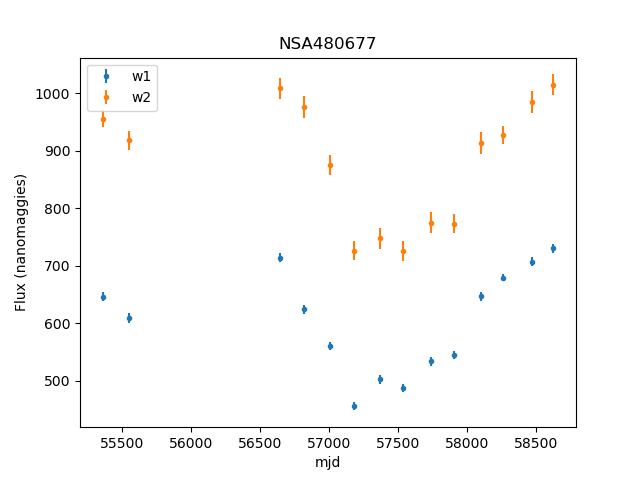}{0.32\textwidth}{NSA480677}}
\gridline{\fig{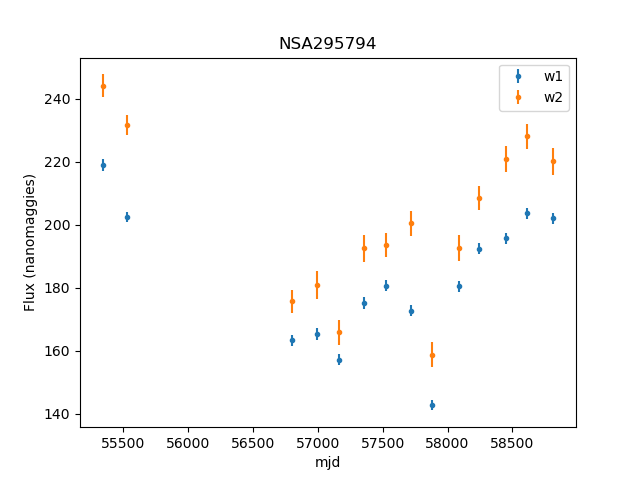}{0.32\textwidth}{NSA295794} \fig{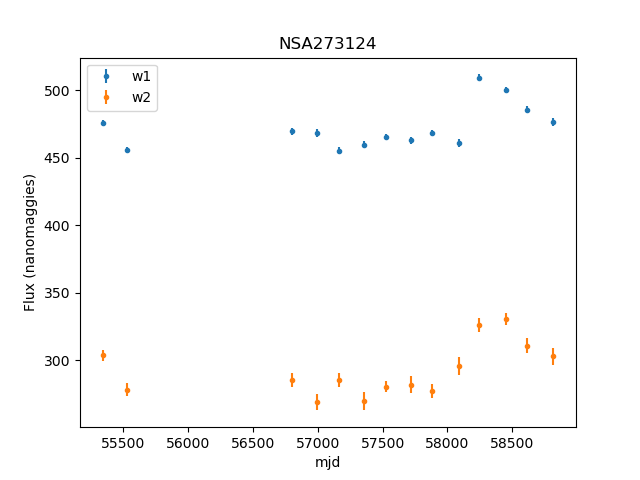}{0.32\textwidth}{NSA273124} 
 \fig{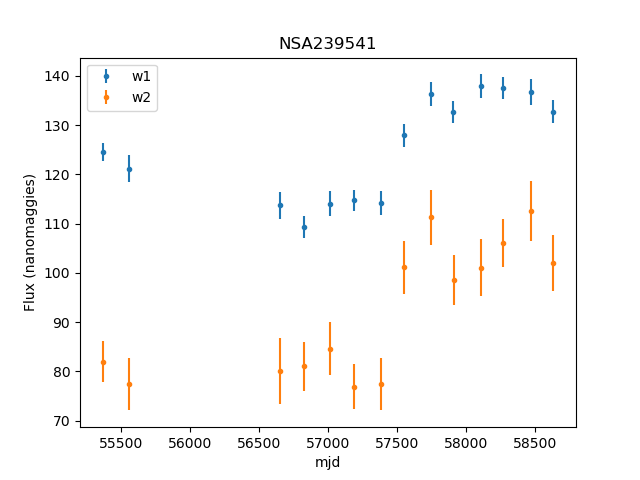}{0.32\textwidth}{NSA239541}}
\caption{Six examples of \textit{WISE} forward modeling light curves which passed the variability criteria and were classified as IMBH candidates.}
\label{fig:lcs_wise}
\end{figure*}

\begin{figure*}
\gridline{\fig{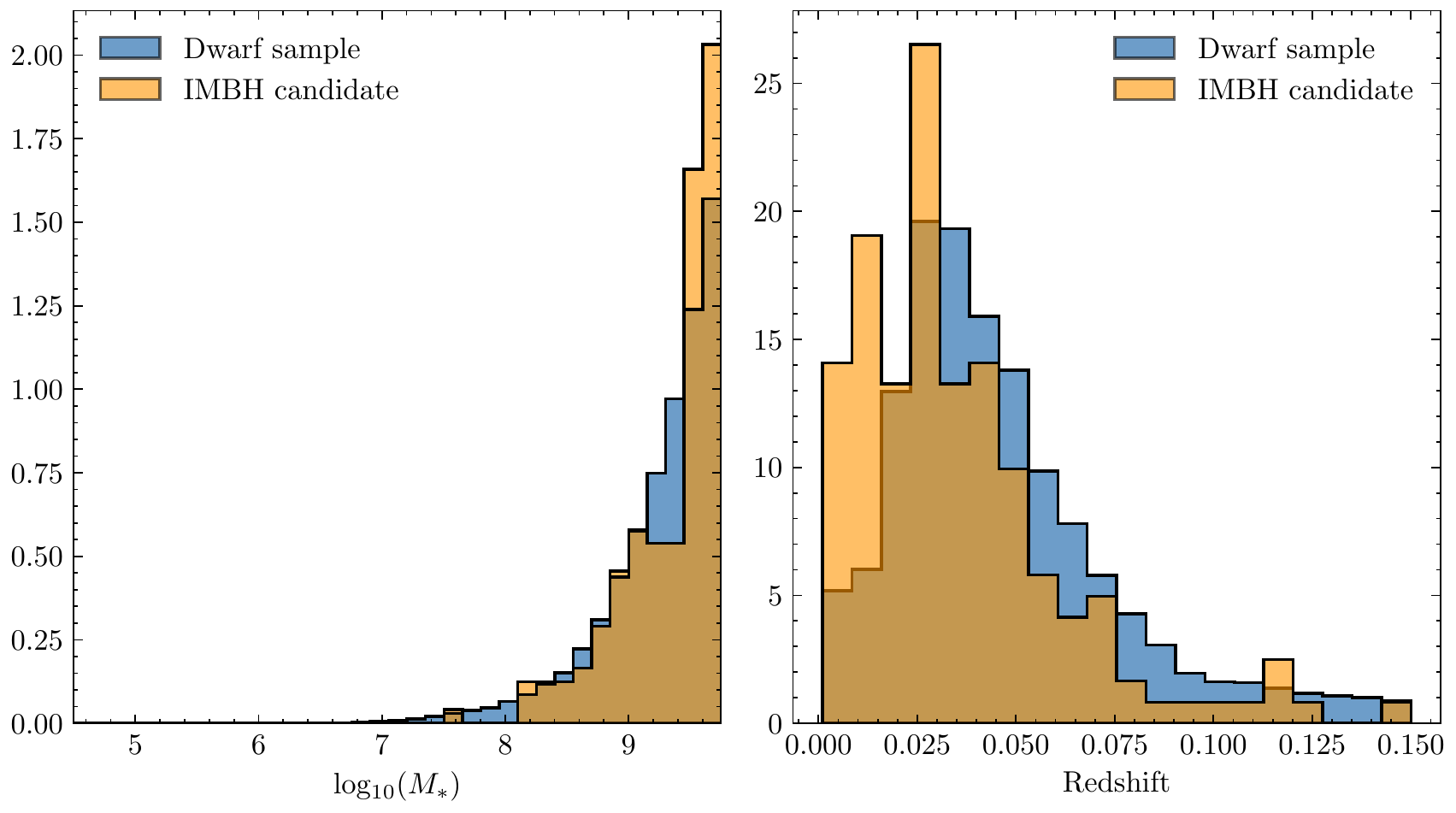}{0.85\textwidth}{}}
\caption{Normalized galaxy mass and redshift distributions of \textit{WISE} IMBH candidates shown in green and the parent dwarf galaxy sample shown in blue.  Redshifts and host galaxy mass measurements are derived from the NSA.}
\label{fig:galstat_wise}
\end{figure*}
\begin{figure*}
\gridline{\fig{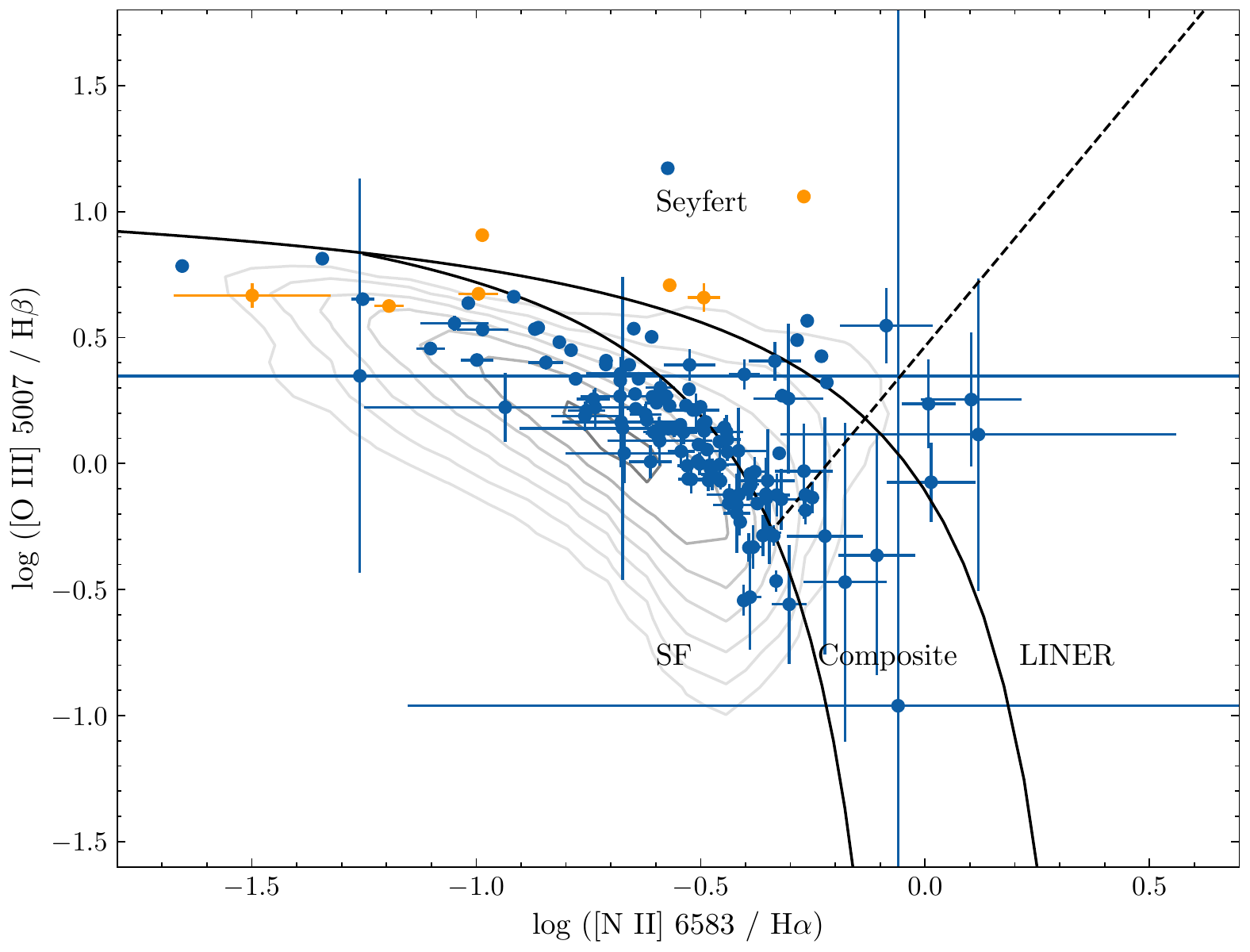}{0.55\textwidth}{}}
\caption{BPT diagram showing narrow line ratios derived from \texttt{pPXF} fitting of archival SDSS spectra of the \textit{WISE}--selected IMBH candidates. Orange points show line ratios for IMBH candidates with broad Balmer lines in archival spectra and blue points show those with narrow emission lines only. Grey contours show the population density with log scaling of the entire parent dwarf galaxy sample for line ratios derived from the NASA--Sloan Atlas. Classification regions are labeled in black text.}
\label{fig:bpt_wise}
\end{figure*}
\begin{figure*}
\gridline{\fig{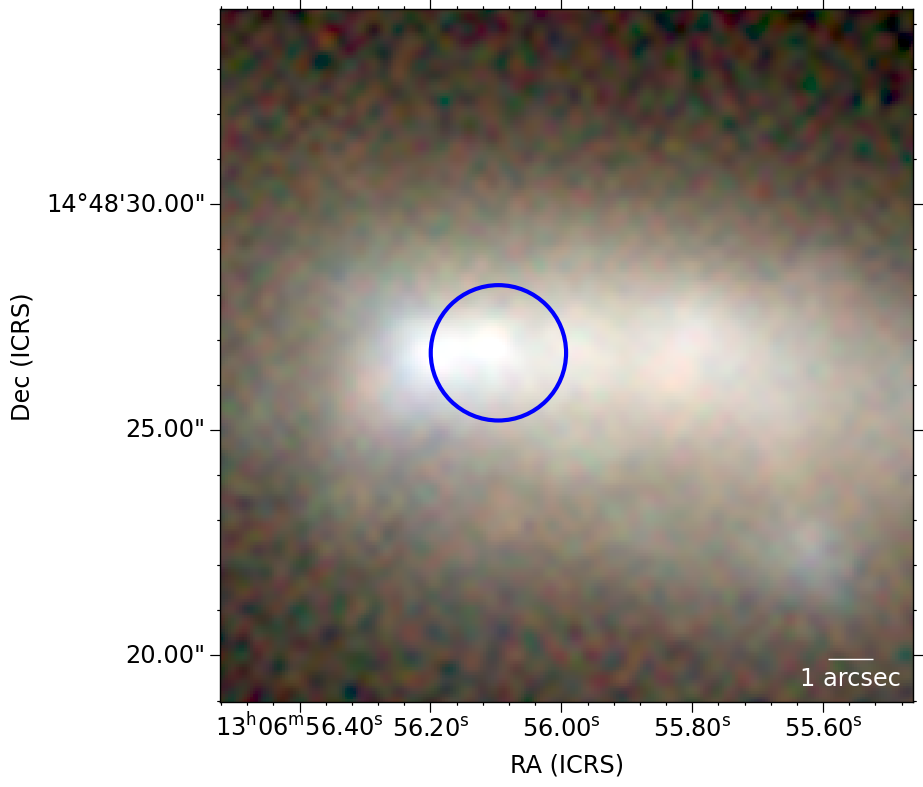}{0.32\textwidth}{(a) NSA370005}\fig{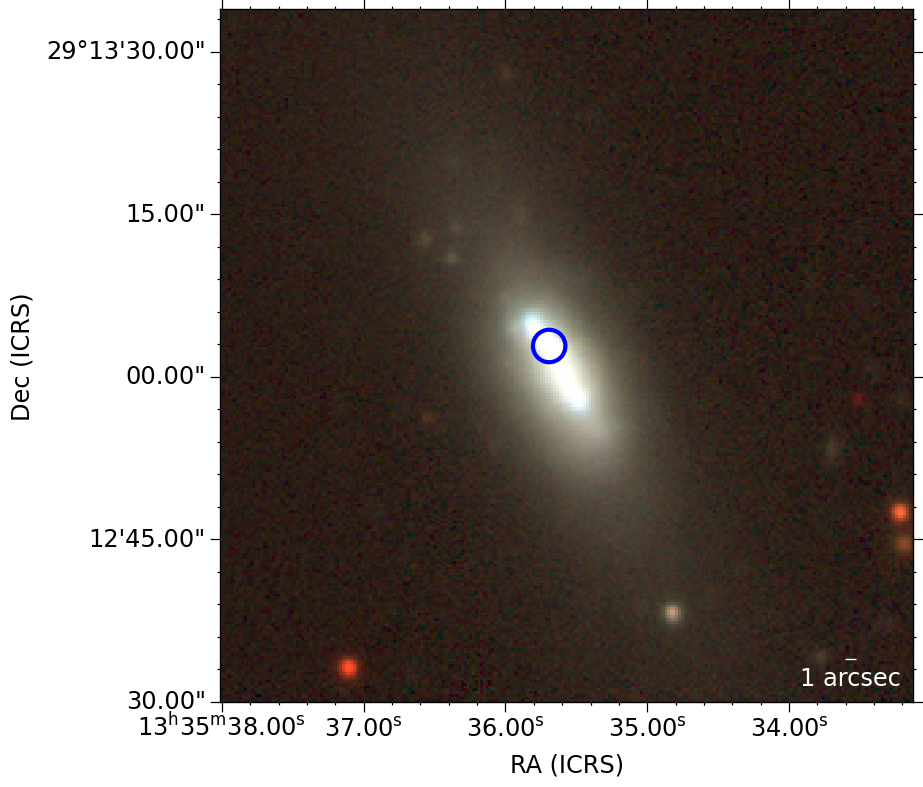}{0.32\textwidth}{(b) NSA418763}\fig{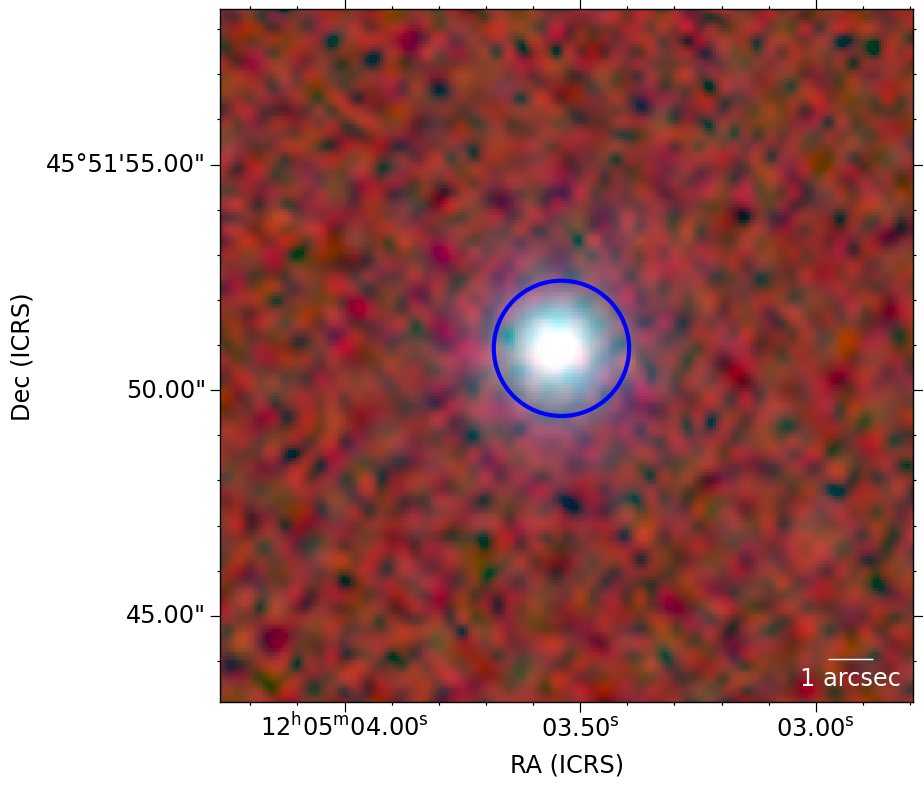}{0.32\textwidth}{(c) NSA612283}}
\gridline{\fig{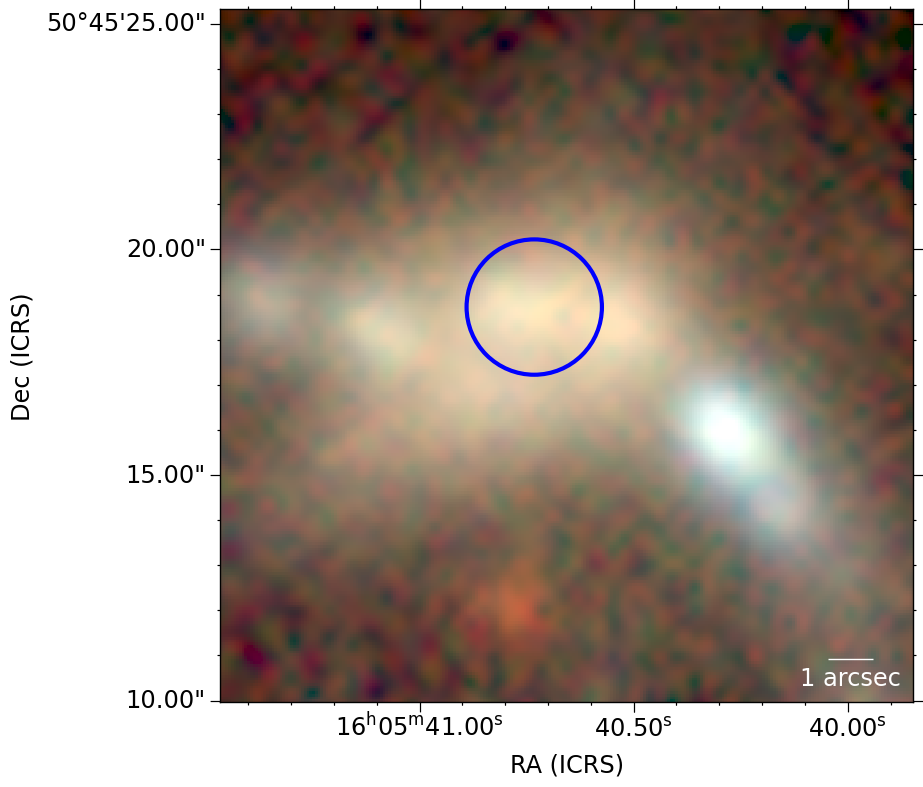}{0.32\textwidth}{(d) NSA97216}\fig{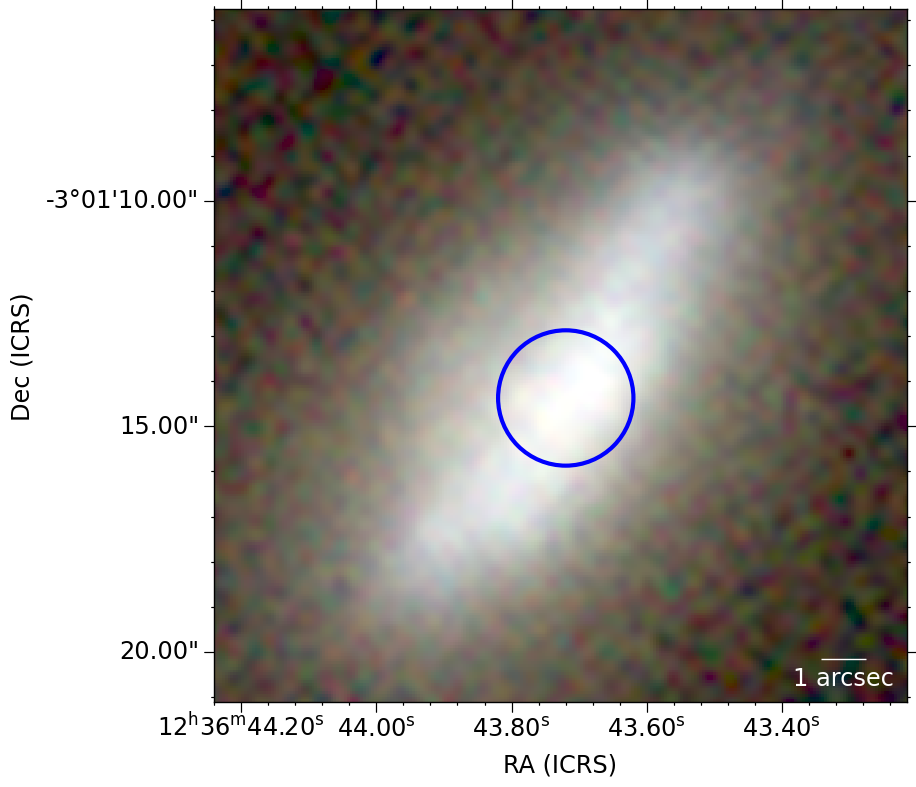}{0.32\textwidth}{(e) NSA19864}\fig{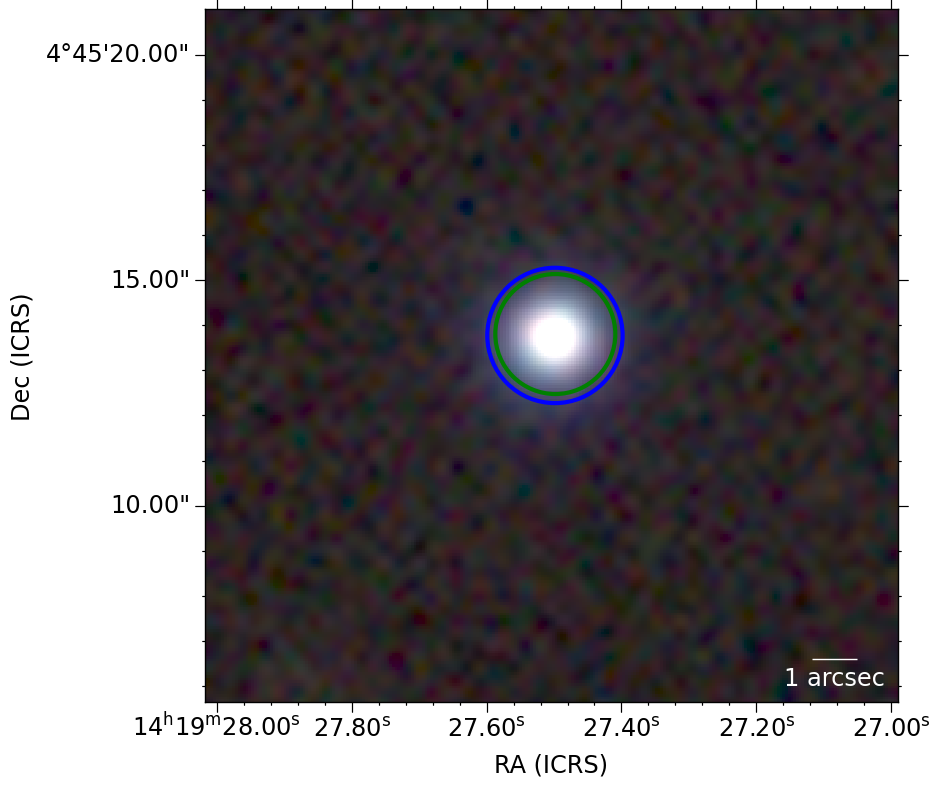}{0.32\textwidth}{(f) NSA87109}}
\caption{Legacy survey cutouts of 3 \textit{WISE} IMBH candidates (a--c) and 2 ZTF IMBH candidates (d--e) with host masses $M_*<10^{8.2}$M$_\odot$ and the candidate with a FIRST radio detection (f, radio position overlaid in green circle). Blue circles show the central position for \textit{WISE} or ZTF forced photometry.} 
\label{fig:decals}
\end{figure*}

As we aimed to search for mid--IR variability from a large sample of dwarf galaxies which may have been too faint to appear in the search of the original All\textit{WISE} catalog by \citet{Secrest2020AGalaxies}, we made use of forward modelled photometry of time--resolved \textit{WISE} coadds \citep{Meisner2018Time-resolvedCoadds} made available through Data Release 9\footnote{https://www.legacysurvey.org/dr9/catalogs/} of the DESI imaging Legacy Survey \citep{Dey2019}. This approach was previously implemented by \citet{Lang2014} to produce forced photometry of 400 million SDSS sources with the deep un\textit{WISE} coadds \citep{Lang2014a,Meisner2017}. \citet{Lang2014} used \texttt{The Tractor} \citep{2016ascl.soft04008L} to use deeper, higher resolution source models from SDSS to produce photometry of blended and faint objects in the un\textit{WISE} coadds. They were able to report fluxes and uncertainties from $3\sigma$ and $4\sigma$ detections which were not included in the original \textit{WISE} catalog. More recently, they implemented this technique to produce time--resolved \textit{WISE} photometry \citep{Meisner2018Time-resolvedCoadds}. As \textit{WISE} revisits each field at a $\sim$6 month cadence, with an increased cadence towards the poles, and takes $\sim10$ exposures each visit, they coadded the exposures from each visit to produce forced photometry on each coadd. This therefore provided light curves with approximately 15 fluxes measured over $\sim8$ years. 

We crossmatched the SDSS galaxy positions of our dwarf galaxy sample to the un\textit{WISE} source catalog and pulled the single epoch \textit{WISE} photometry for the closest un\textit{WISE} source. Some light curves showed an oscillation behaviour when multiple sources were contained within the large \textit{WISE} PSF and the \textit{WISE} flux was distributed across the multiple sources differently in each epoch. To overcome this, we combined the total flux of the sources within a radius of 3 \textit{WISE} pixels (3x2.75") to ensure that all dwarf galaxy variability was captured within the combined fluxes. We removed light curves where an erroneously high flux ( $f-\langle f \rangle>5\sigma$) in both W1 and W2 bands produced a $r$ and $\chi^2/N$ above the cutoffs. We also removed light curves where source confusion within the \textit{WISE} PSF size still resulted in an oscillating behaviour using the following criteria. If the summed difference between every adjacent W1 flux measurement was greater than twice the summed difference between every W1 flux offset by two epochs, the light curve was flagged as bad quality. We produced \textit{WISE} light curves of the AGN control sample with the same procedure. 

For each source we calculated the Pearson correlation coefficient $r$ between the W1 and W2 bands and the $\chi^2/N$ for each band. The distribution of these statistics for the two samples is shown in Figure \ref{fig:varstat_wise}. We required that dwarf galaxies have $r>0.75$ and $\chi^2/N>1.0$ in W1 and W2 to be considered a variable AGN candidate. 

Of the 79,879 dwarf galaxies for which \textit{WISE} single epoch forced photometry was available, 124 were removed due to light curve quality flags. Of the remaining 79,755 light curves, we found that 165 had fluxes which met our variability criteria. One dwarf galaxy, NSA253466, was detected by our variability criteria due to the well--studied Type IIn supernova SN 2010jl \citep{Stoll2011SNGalaxy} which has had detailed follow--up in the near--IR, X--ray and radio \citep[e.g.][]{Fransson2014High-densityDays,Chandra2015X-RAY2010jl} showing interaction with the dense circumstellar medium. We removed this object from our AGN candidate sample. We removed 3 other objects with supernovae visible in both WISE and ZTF data: NSA559938 (ZTF18aamftst: an SNIIn), NSA143427 (ZTF18acwyvet, an SNIIL), and NSA230430 (ZTF20aaupkac: an SNIa). We removed 14 other galaxies with light curves showing single flares, often with color changes characteristic of SNe: NSA20892, NSA32356, NSA143207, NSA236644, NSA250558, NSA253466, NSA274965, NSA340533, NSA355173, NSA379733, NSA475418, NSA502699, NSA528212, NSA548379. The properties of these supernova host galaxies are summarized in Table \ref{table:wiseSN}.  

The properties of the remaining final sample of 148 AGN candidates (corresponding to $0.19\pm0.02$\% of the dwarf galaxy sample) are summarized in Table \ref{table:wisecands} and 6 examples of variable \textit{WISE} light curves are shown in Figure \ref{fig:lcs_wise}. 

\section{Properties of the \textit{WISE}--selected variable IMBH candidates}
The distributions of the redshifts and host stellar masses of the mid--IR variable dwarf galaxies are shown in Figure \ref{fig:galstat_wise}. We exclude NSA64525 from the histogram because the estimated mass provided by the NASA--Sloan Atlas ($10^{5.39}M_\odot$) is inconsistent with the stellar dispersion velocity of $193\pm8$ km/s measured by the SDSS spectroscopic pipeline. This dispersion velocity indicates that the BH itself likely has a mass of $\sim10^7M_\odot$ according to the $M-\sigma_*$ relation \citep{Kormendy2013CoevolutionGalaxies} and the host galaxy therefore has mass comparable to $\sim10^{9.5}M_\odot$ based on the $M_{\text{BH}}-M_*$ relation \citep{Schutte2019TheNuclei}. The mid--IR variability selection finds a higher fraction of variable AGN at low redshifts compared to the overall dwarf galaxy sample and shows a slight preference for higher mass galaxies.

The classification of the mid--IR IMBH candidates on the [O\,{\sc iii}] $\lambda$5007/H$\beta$ -- [N\,{\sc ii}] $\lambda$6583 /H$\alpha$ BPT diagram is shown in Figure \ref{fig:bpt_wise}. 100 objects (69\%) are classified as star forming while 32 objects (22\%) are in the composite region, 3 (2.1\%) were classified as LINERs and 10 (6.9\%) as Seyferts. 12 of the IMBH candidates (8.1\%) have broad Balmer lines. These objects also had their virial black hole masses estimated using the width of the H$\alpha$ broad lines by \citet{Ho2015ASPECTRA} and \citet{Liu2019ADR7}. These virial masses range between $10^{5.5}M_\odot$ and $10^{8.2}M_\odot$ respectively and are shown in Table \ref{table:wisecands}.

In order to determine how many \textit{WISE}--selected IMBH candidates were also optically variable, we produced ZTF photometry of the remainder of the 148 dwarf host galaxies which were not included in the original ZTF search. We found that 15 out of 148 (10\%) met the optical variability criteria for ZTF outlined in Section 3.5. 7 of the 15 AGN which were variable at both wavelengths had visible broad lines in their spectra, and amongst these broad line AGN, the virial masses ranged between $M_{\text{BH}}=10^{6.3}M_\odot$ and $M_{\text{BH}}=10^{8.2}M_\odot$. Only two of these sources (NSA164884 and NSA451469) were found in the original ZTF sample, due to the smaller sample of dwarf galaxies for which we produced optical photometry. The AGN with both mid--IR and optical variability are indicated in the last column of Table \ref{table:wisecands}.

NSA451469 had also been found in PTF by \citet{Baldassare2020AFactory}. We crossmatched our mid--IR selected IMBH candidates to the active dwarf galaxies from the \citet{Secrest2020AGalaxies} mid--IR variability search, the \citet{Baldassare2018,Baldassare2020AFactory} optical variability searches, the \citet{Molina2021AEvents} [Fe\,X] $\lambda$6374 coronal line emission search, the \citet{Mezcua2020HiddenAGN} IFU spectroscopy search, the \citet{Mezcua2018Intermediate-massSurvey} Chandra X-ray search, the \citet{Latimer2021AGalaxies} mid--IR color selection box search and the \citet{Reines2013DWARFHOLES} optical emission line search. Only 2 objects had been found in previous AGN searches.

One object, NSA638093, had previously been found in the \textit{WISE} variability search by \citet{Secrest2020AGalaxies} and in the mid--IR color selection box by \citet{Latimer2021AGalaxies} (object number 11, listed with NSAID 151888 from version \texttt{v$_{1\textunderscore 1\textunderscore 2}$} of the NSA), where they used {\it Chandra} to find X-ray emission which may have been consistent with X-ray binaries instead of AGN activity. NSA386591 appears in the \citet{Reines2013DWARFHOLES} spectroscopic search for BPT AGN and Composites and the \citet{Latimer2021AGalaxies} mid--IR color selection search (ID number 6) where it was found to have an X-ray luminosity consistent with AGN activity and too large to be produced by X-ray binaries. 

We undertook a search for radio emission from the IMBH candidates in the Karl G. Jansky Very Large Array Sky Survey \citep[VLASS;][]{Lacy2020TheDesign}. This survey covers a total of 33,885 deg$^2$ in the 2--4 GHz range with an angular resolution of $\sim 2".5$ and will obtain a coadd $1\sigma$ sensitivity of 1 $\mu$Jy/beam by survey end in 2024. We searched for crossmatches within $10"$ in Table 2 of the VLASS Epoch 1 Quick Look Catalogue which contains $\approx 700,000$ compact radio sources with $>1$ mJy/beam detections associated with mid--IR hosts from the un\textit{WISE} catalog \citep{Gordon2020AIdentifications}. We found that one IMBH candidate, NSA87109, had a corresponding radio point source at a separation of 0.0038" from the optical NASA--Sloan Atlas galaxy position with a flux of $17.55\pm0.27$ Jy. This object is a known BL Lac which has also been detected in 0.1--2.4 keV X-rays \citep{Massaro2009Roma-BZCAT:Blazars}.

We show cutouts from the DESI Legacy Imaging Surveys \citep{Dey2019} of 3 \textit{WISE} candidates and 2 ZTF candidates with host masses $M_*<10^{8.2}M_\odot$ and the object with a VLASS radio detection (NSA87109, with stellar mass $\log_{10}M_*=9.66$) in Figure \ref{fig:decals}. The radio source and one other low mass IMBH candidate are in compact, blue galaxies while the other 4 low mass IMBH candidates reside in galaxies with complex morphologies and multiple stellar overdensities.

\section{Discussion}

The number of dwarf galaxies which were variable in \textit{WISE} corresponded to a $0.19\pm0.02$\% variability fraction, while the optical variability fraction that we find from the ZTF AGN candidates is $0.17\pm0.03$\%. Our results therefore suggest that the two methods are similarly effective for identifying IMBH candidates with the current sensitivities and baselines available for mid--IR and optical photometry of dwarf galaxies. This, however, will change with the improved optical sensitivities and baselines offered by the LSST survey over the next decade.

Our mid--IR active fraction was larger but within the uncertainty range of the active fraction found by \citet{Secrest2020AGalaxies} ($0.09^{+0.20}_{-0.07}$\%) in a much smaller sample of 2197 dwarf galaxies with the main \textit{WISE} photometry catalog. The optical variability fraction that we find is also consistent with the active fraction of $0.15\pm0.07$\% found for dwarf galaxies of the same mass in PTF \citep{Baldassare2020AFactory}. 

The 90\% fraction of \textit{WISE}-selected AGN candidates which are variable in the mid-IR but not the optical likely arises from a combination of line-of-sight obscuration of nuclear optical emission due to the dusty torus and global obscuration from the host galaxy. Obscuration of nuclear optical emission from AGN with detectable mid--IR signatures has been observed for many Seyfert 2 galaxies \citep[e.g.][]{Goulding2011Searching0.1,Annuar2015NuSTAR5643,Ricci2016NuSTAR6286,Annuar2017NuSTAR}. It is possible that the majority of our mid--IR variable AGN in dwarf galaxies are Seyfert 2s with obscured optical variability which are also not picked up by the BPT diagram due to the effects of their lower masses on the optical emission line ratios. Line-of-sight obscuration of optical emission of \textit{WISE} candidates is supported by the fact that 7 of the 12 \textit{WISE}--selected IMBH candidates with bright BLRs were variable in both the mid--IR and the optical.

We note that only 5 of the 152 dwarf galaxies in common with the \citet{Baldassare2020AFactory} AGN candidate sample from PTF were variable in ZTF. This may be because of the use of differing statistical criteria for variability classification. It may also be the case that the longer 7 year baseline of the combined PTF and iPTF light curves in \citet{Baldassare2020AFactory} improved their sensitivity to AGN which vary on longer timescales compared to shorter timescales. They indeed find that longer baseline data has a much higher detectable variability fraction, increasing by a factor of 4 from 0.25\% for light curves with $<2$ year baselines to 1\% for light curves with $>2$ year baselines. By comparison, our procedure of using deep references and stacking to detect fainter variability in more evenly--sampled, $\sim3$ year baseline light curves may make us more sensitive to variability on month to year long timescales, but may miss AGN with flux changes over longer 2--7 year timescales. An alternative explanation may be that a large fraction of low mass AGN are state--changing AGN \citep[e.g.][]{Frederick2019ALINERs} which can switch their optical variability on and off over the decade--long timescale of the PTF and ZTF surveys. Indeed, \citet{Martinez-Palomera2020IntroducingSurvey} found that the majority of IMBH candidates with optical variability on hourly to daily timescales from the SIBLING survey were not variable when observed again the following year.

81\% of the ZTF--selected IMBH candidates were star--forming on the BPT diagram and only 7 have been identified as AGN via their Balmer broad lines. Only 7 had been identified in previous dwarf galaxy AGN searches and these were all via their optical variability in SDSS or PTF. The non--AGN spectroscopic classifications of the majority of the sources indicates that optical variability selection can find AGN in dwarf galaxies which would be missed by other selection strategies. 

Similarly, 100 (69\%) of \textit{WISE}--selected IMBH candidates were star--forming on the BPT diagram and therefore would have been missed by classic spectroscopic selection methods. 12 (8.1\%) can be identified as AGN due to the presence of broad lines, 1 (0.63\%) could have been found via radio emission alone, another 2 (1.27\%) via the mid--IR selection box and 1 (0.63\%) via X-ray emission. We therefore see that $\sim70$\% of candidates from mid--IR variability selection could likely not have been found through other selection techniques. A higher fraction of the \textit{WISE}--selected candidates are BPT--AGN compared to optically--selected candidates, perhaps indicating that the AGN emission lines of galaxies with mid--IR variability are less likely to be diluted by star formation. 

Both the ZTF--selected and \textit{WISE}--selected AGN candidate host galaxies tend to have higher masses and lower redshifts compared to the overall distributions of the host galaxy sample, likely due to the higher luminosities of these AGN. However, our selection method is still capable of detecting AGN variability from dwarf galaxies with redshifts up to $z\sim0.15$ and stellar masses down to $M_*=10^{7.52}M_\odot$. The virial masses for the AGN candidates with broad lines go down to $M_{\text{BH}}=10^{5.485}M_\odot$ and in most cases are lower than the BH masses estimated from the $M_\text{BH}-M_*$ relation, indicating that our search may have found MBHs which are undermassive for their hosts. We therefore conclude that our variability--selection approach is useful for selecting AGN which can populate the poorly sampled lower end of the $M_{\text{BH}}-\sigma_*$ relation. Future work should take high resolution spectra of our IMBH candidates for fitting of the velocity dispersion from the stellar absorption lines to provide another independent estimate of the BH mass.

The discovery of 36 nuclear supernova candidates shows the usefulness of applying simple statistics to ZTF forced photometry of large galaxy samples to find supernovae candidates in dwarf galaxies. There are many motivations to study rates of supernovae in dwarfs such as explaining the increase in the rate ratio of superluminous supernovae to core collapse supernovae in low mass galaxies and whether the increased metallicity or specific star formation rates in dwarfs are the driving factor for this trend \citep{Taggart2021Core-collapseGalaxies}. It has also been found that the emerging class of fast blue optical transients like AT2018cow are preferentially hosted by dwarf galaxies \citep{Perley2021Real-timeEjecta}. Our finding that 0.14\% of dwarf galaxies contained nuclear supernovae during ZTF Phase--I (corresponding to 0.05\% per year) and that most SN were within redshifts of $0.02<z<0.055$ may provide insights into these questions.

\section{Conclusions}
In this paper we have presented a search for IMBH candidates by looking for variable AGN in dwarf galaxies of stellar mass $M_*<10^{9.75}M_\odot$ in the optical and mid--IR. We applied a new ZTF forced photometry pipeline to produce deep, high quality reference images for image subtraction and made g- and r- band light curves of 25,714 dwarf galaxies. These light curves were stacked in a range of time bins to improve sensitivity to faint variability. We applied statistical cutoffs to find significant and correlated variability between the two bands and found 36 supernova candidates and 44 AGN candidates. The supernova fraction was $0.05\pm0.01$\% year$^{-1}$ and the optically variable AGN fraction was $0.17\pm0.03$\%. 

To search for mid--IR variability we used \texttt{Tractor} forward modelled photometry of time--resolved \textit{WISE} coadds. We found 148 dwarf galaxies with significant and correlated variability in the W1 and W2 bands after removing 14 supernovae. The mid--IR variable AGN fraction was $0.19\pm0.02$\%. 

We found that 81\% of our ZTF AGN candidates would have been missed with classical spectroscopic classification on the BPT diagram. Of our \textit{WISE} candidates, 69\% would have been missed with spectroscopic classification and only 4 would have been detected via radio or X-ray detection or the mid--IR color selection box. While our candidates were slightly biased to low redshifts and high galaxy masses compared to the parent dwarf galaxy sample, they were effective in identifying AGN with virial masses as low as $M_{\text{BH}}=10^{5.485}M_\odot$ and in dwarf galaxies with stellar masses $10^{7.5}<M_*<10^{8.5}M_\odot$. We therefore conclude, in accordance with previous variability searches, that optical and mid--IR variability selection is effective for finding low mass AGN in dwarf galaxies which would be missed by other spectroscopic selection techniques. 

After checking the ZTF photometry of the 152 dwarf galaxies in common between our parent sample and the variability--selected AGN candidates from PTF \citep{Baldassare2020AFactory} we found that only 5 continue to show variability in ZTF. The lack of variability in our ZTF light curves may be due to the different baselines and sensitivities of the two search strategies and the differing methods for finding AGN--like variability.

With more detailed imaging and spectroscopic analysis, our variability--selected AGN candidates could help to populate the sparsely sampled end of the $M_*-\sigma$ relation and provide insights into black hole seed formation mechanisms, dwarf galaxy--black hole co--evolution and the accretion states of low mass AGN. Future work on these candidates will include forward modeling of ZTF and DECam images to determine their positions relative to their host galaxy nuclei and determine the off-nuclear fraction. The potential for forward modeling of ZTF images to determine the position of a variable point source relative to its host galaxy was demonstrated in \citet{Ward2021AGNsFacility} for recoiling SMBH candidates and will provide a way to confirm the positions of IMBH candidates without X-ray or radio detections. 

These candidates are just the tip of the iceberg in the search for optically variable AGN in dwarf galaxies, which will be greatly enhanced by the capabilities of the Legacy Survey of Space and Time (LSST) at Vera C. Rubin Observatory \citep{Ivezic2019LSST:Products} over the next decade. The capacity of LSST to find fainter and more distant IMBHs in dwarf galaxies via their variability will tighten the constraints we can place on black hole seeding channels and the efficiency of massive BH growth.

\section{Acknowledgements}
We thank the anonymous referee for their very helpful feedback which has helped us to substantially improve the manuscript. The authors would also like to thank Yuhan Yao for her useful comments.

Based on observations obtained with the Samuel Oschin Telescope 48-inch and the 60-inch Telescope at the Palomar Observatory as part of the Zwicky Transient Facility project. ZTF is supported by the National Science Foundation under Grant No. AST-2034437 and a collaboration including Caltech, IPAC, the Weizmann Institute for Science, the Oskar Klein Center at Stockholm University, the University of Maryland, Deutsches Elektronen-Synchrotron and Humboldt University, the TANGO Consortium of Taiwan, the University of Wisconsin at Milwaukee, Trinity College Dublin, Lawrence Livermore National Laboratories, and IN2P3, France. Operations are conducted by COO, IPAC, and UW. The ZTF forced-photometry service was funded under the Heising-Simons Foundation grant 12540303 (PI: Graham). ECK acknowledges support from the G.R.E.A.T research environment funded by {\em Vetenskapsr\aa det}, the Swedish Research Council, under project number 2016-06012, and support from The Wenner-Gren Foundations. MMK acknowledges generous support from the David and Lucille Packard Foundation. This work was supported by the GROWTH project funded by the National Science Foundation under Grant No 1545949. 

This research used resources of the National Energy Research Scientific Computing Center, a DOE Office of Science User Facility supported by the Office of Science of the U.S. Department of Energy under Contract No. DE-AC02-05CH11231. P.E.N. acknowledges support from the DOE under grant DE-AC02-05CH11231, Analytical Modeling for Extreme-Scale Computing Environments.

This project used data obtained with the Dark Energy Camera (DECam), which was constructed by the Dark Energy Survey (DES) collaboration. Funding for the DES Projects has been provided by the U.S. Department of Energy, the U.S. National Science Foundation, the Ministry of Science and Education of Spain, the Science and Technology Facilities Council of the United Kingdom, the Higher Education Funding Council for England, the National Center for Supercomputing Applications at the University of Illinois at Urbana-Champaign, the Kavli Institute of Cosmological Physics at the University of Chicago, Center for Cosmology and Astro-Particle Physics at the Ohio State University, the Mitchell Institute for Fundamental Physics and Astronomy at Texas A\&M University, Financiadora de Estudos e Projetos, Fundacao Carlos Chagas Filho de Amparo, Financiadora de Estudos e Projetos, Fundacao Carlos Chagas Filho de Amparo a Pesquisa do Estado do Rio de Janeiro, Conselho Nacional de Desenvolvimento Cientifico e Tecnologico and the Ministerio da Ciencia, Tecnologia e Inovacao, the Deutsche Forschungsgemeinschaft and the Collaborating Institutions in the Dark Energy Survey. The Collaborating Institutions are Argonne National Laboratory, the University of California at Santa Cruz, the University of Cambridge, Centro de Investigaciones Energeticas, Medioambientales y Tecnologicas-Madrid, the University of Chicago, University College London, the DES-Brazil Consortium, the University of Edinburgh, the Eidgenossische Technische Hochschule (ETH) Zurich, Fermi National Accelerator Laboratory, the University of Illinois at Urbana-Champaign, the Institut de Ciencies de l'Espai (IEEC/CSIC), the Institut de Fisica d'Altes Energies, Lawrence Berkeley National Laboratory, the Ludwig-Maximilians Universitat Munchen and the associated Excellence Cluster Universe, the University of Michigan, the National Optical Astronomy Observatory, the University of Nottingham, the Ohio State University, the University of Pennsylvania, the University of Portsmouth, SLAC National Accelerator Laboratory, Stanford University, the University of Sussex, and Texas A\&M University.

\software{
Ampel \citep{Nordin2019},
Astromatic (\url{https://www.astromatic.net/}),
Astropy \citep{Robitaille2013Astropy:Astronomy,Price-Whelan2018ThePackage}, 
catsHTM \citep{Soumagnac2018},
extcats (\url{github.com/MatteoGiomi/extcats}),
GROWTH Marshal \citep{Kasliwal_2019},
Hotpants (\url{https://github.com/acbecker/hotpants}),
The Tractor \citep{2016ascl.soft04008L}.
}

\startlongtable
\begin{deluxetable*}{rcccccccc}
\tabletypesize{\scriptsize}
\tablecolumns{10}
\tablewidth{0pt}
\tablecaption{Properties of the 14 SN candidates from  \textit{WISE}. \label{table:wiseSN}}
\tablehead{
\colhead{NSA ID} & \colhead{RA (hms) } & \colhead{Dec (dms)} & \colhead{z} &\colhead{log$_{10}$M$_*$($M_\odot$)}&\colhead{log$_{10}$M$_{\text{BH}}$($M_\odot$)}& \colhead{BPT class} & \colhead{BLR?}&\colhead{SN Class} }
\startdata
20892&12:59:05.105&-03:12:14.257&$0.0101$&$8.19$&$5.31\pm0.68$&SF&$\times$&None\\
32356&00:10:05.515&-01:05:39.795&$0.0379$&$9.43$&$6.85\pm0.68$&SF&$\times$&None\\
143207&16:28:31.291&41:37:12.833&$0.0331$&$9.59$&$7.05\pm0.68$&SF&$\times$&None\\
143427&16:32:03.899&39:36:31.289&$0.0281$&$9.27$&$6.66\pm0.68$&SF&$\times$&SNIIL\\
230430&07:51:08.935&22:44:27.419&$0.0288$&$9.25$&$6.63\pm0.68$&-&$\times$&SNIa\\
236644&11:43:039.35&09:50:18.863&$0.0202$&$8.77$&$6.03\pm0.68$&SF&$\times$&None\\
250558&15:31:51.855&37:24:047.41&$0.0299$&$9.26$&$6.65\pm0.68$&-&$\times$&None\\
253466&09:42:53.245&09:29:36.066&$0.0106$&$9.06$&$6.4\pm0.68$&-&$\times$&None\\
274965&13:08:19.122&43:45:25.483&$0.0365$&$9.57$&$7.03\pm0.68$&-&$\times$&None\\
355173&15:28:40.931&10:33:28.095&$0.0887$&$9.71$&$7.19\pm0.68$&SF&\checkmark&None\\
379733&13:32:18.328&07:56:26.052&$0.0236$&$9.57$&$7.03\pm0.68$&SF&$\times$&None\\
502699&10:08:07.779&19:17:57.808&$0.0335$&$9.5$&$6.94\pm0.68$&Composite&$\times$&None\\
548379&13:20:53.681&21:55:10.518&$0.0224$&$8.63$&$5.86\pm0.68$&-&$\times$&None\\
559938&14:31:14.719&17:11:35.152&$0.0378$&$9.38$&$6.79\pm0.68$&SF&$\times$&SNIIn\\
\enddata
\vspace{0.1cm}
\tablecomments{Properties of the 14 SN candidates found in forward modeled \textit{WISE} light curves. The IDs, positions, redshifts and host galaxy stellar masses are those from the NSA catalog version \texttt{v$_{1\textunderscore0\textunderscore1}$}. In the log$_{10}$M$_{BH}$ column we show the estimated black hole mass based on the $M_*-M_{BH}$ from \citet{Schutte2019TheNuclei} which has a scatter of 0.68 dex. The presence of Balmer broad lines is indicated in the BLR column. The last column shows the spectroscopic classification made published by ZTF on the Transient Name Server.}
\end{deluxetable*}

\startlongtable
\begin{deluxetable*}{rccccccccc}
\tabletypesize{\scriptsize}
\tablecolumns{11}
\tablewidth{0pt}
\tablecaption{Properties of the 148 \textit{WISE}-selected IMBH candidates \label{table:wisecands}}
\tablehead{
\colhead{NSA ID} & \colhead{RA (hms) } & \colhead{Dec (dms)} & \colhead{z} &\colhead{log$_{10}$M$_*$($M_\odot$)}&\colhead{log$_{10}$M$_{\text{BH}}$($M_\odot$)}& \colhead{BPT class} & \colhead{BLR?}&\colhead{log$_{10}M_{\text{BH,vir}}(M_\odot$)} &\colhead{Optical var.?}}
\startdata
3045&10:39:44.299&00:51:28.625&$0.0253$&$8.88$&$6.17\pm0.68$&SF&$\times$&-&$\times$\\
12740&14:06:30.104&00:19:39.525&$0.1063$&$8.93$&$6.24\pm0.68$&SF&\checkmark&6.436&$\times$\\
34102&00:45:000.51&00:47:23.578&$0.0568$&$9.63$&$7.1\pm0.68$&SF&$\times$&-&$\times$\\
47457&08:03:38.042&43:20:034.99&$0.0153$&$9.35$&$6.75\pm0.68$&SF&$\times$&-&$\times$\\
50093&08:38:19.706&51:31:52.998&$0.0166$&$9.03$&$6.35\pm0.68$&SF&$\times$&-&$\times$\\
50870&08:48:04.539&52:14:09.051&$0.0403$&$9.6$&$7.07\pm0.68$&Seyfert&$\times$&-&$\times$\\
54469&03:19:26.057&-6:07:015.98&$0.0076$&$9.45$&$6.87\pm0.68$&SF&$\times$&-&$\times$\\
64525&10:11:03.771&02:31:45.452&$0.1211$&$5.39$&$1.84\pm0.68$&LINER&$\times$&-&$\times$\\
71121&13:19:40.797&03:24:33.828&$0.0218$&$8.27$&$5.42\pm0.68$&SF&$\times$&-&$\times$\\
73327&14:10:052.96&01:22:10.396&$0.026$&$9.73$&$7.23\pm0.68$&SF&$\times$&-&$\times$\\
74770&14:58:54.268&01:59:027.33&$0.0298$&$9.49$&$6.93\pm0.68$&SF&$\times$&-&$\times$\\
75025&15:05:07.305&01:59:51.121&$0.008$&$8.73$&$5.99\pm0.68$&-&$\times$&-&$\times$\\
81096&08:41:34.335&02:11:19.877&$0.0287$&$8.93$&$6.23\pm0.68$&SF&$\times$&-&$\times$\\
87109&14:19:27.498&04:45:13.802&$0.1434$&$9.66$&$7.13\pm0.68$&LINER&$\times$&-&\checkmark\\
93256&14:02:059.68&61:45:03.942&$0.0054$&$9.33$&$6.73\pm0.68$&Composite&$\times$&-&$\times$\\
95784&15:40:37.104&58:15:035.93&$0.0501$&$9.64$&$7.11\pm0.68$&SF&$\times$&-&$\times$\\
98016&16:06:33.607&48:39:37.056&$0.0437$&$9.48$&$6.91\pm0.68$&Composite&$\times$&-&$\times$\\
102352&20:58:22.143&-6:50:04.353&$0.0738$&$9.72$&$7.21\pm0.68$&Seyfert&\checkmark&6.432&\checkmark\\
106103&00:01:11.154&-10:01:55.702&$0.0489$&$9.65$&$7.12\pm0.68$&Seyfert&\checkmark&6.372&$\times$\\
107142&00:23:55.671&-9:24:22.513&$0.0522$&$9.57$&$7.03\pm0.68$&Composite&$\times$&-&$\times$\\
111190&01:52:32.835&-8:25:058.66&$0.0176$&$9.45$&$6.88\pm0.68$&SF&$\times$&-&$\times$\\
112816&23:09:22.927&01:00:002.16&$0.0153$&$9.14$&$6.49\pm0.68$&SF&$\times$&-&$\times$\\
120721&21:30:046.71&11:12:54.805&$0.0475$&$9.62$&$7.09\pm0.68$&SF&$\times$&-&$\times$\\
124852&23:26:18.717&14:11:50.593&$0.0417$&$9.49$&$6.93\pm0.68$&Composite&$\times$&-&$\times$\\
128577&07:47:55.395&34:02:12.144&$0.0159$&$9.1$&$6.45\pm0.68$&SF&$\times$&-&$\times$\\
129617&08:12:26.587&39:32:19.995&$0.0326$&$9.74$&$7.24\pm0.68$&SF&$\times$&-&$\times$\\
130265&08:32:24.255&43:06:07.807&$0.0247$&$9.12$&$6.47\pm0.68$&SF&$\times$&-&$\times$\\
135650&12:40:057.29&63:31:010.47&$0.0083$&$9.67$&$7.14\pm0.68$&Composite&$\times$&-&$\times$\\
143517&16:31:24.055&41:22:44.657&$0.0309$&$9.47$&$6.91\pm0.68$&SF&$\times$&-&$\times$\\
152094&13:16:059.37&03:53:19.809&$0.0454$&$9.71$&$7.2\pm0.68$&Seyfert&$\times$&-&$\times$\\
155198&08:17:15.852&31:39:07.139&$0.0452$&$9.72$&$7.21\pm0.68$&Composite&$\times$&-&$\times$\\
164884&09:28:01.292&49:18:17.303&$0.1143$&$9.71$&$7.21\pm0.68$&SF&\checkmark&6.961&$\times$\\
175367&09:09:33.556&38:42:38.594&$0.0561$&$9.69$&$7.17\pm0.68$&SF&$\times$&-&$\times$\\
178387&10:39:28.739&56:20:58.148&$0.0748$&$9.61$&$7.08\pm0.68$&Seyfert&\checkmark&6.515&\checkmark\\
180304&12:03:38.324&58:59:11.804&$0.0469$&$9.6$&$7.06\pm0.68$&SF&$\times$&-&$\times$\\
185495&12:05:21.947&50:14:11.506&$0.0341$&$8.8$&$6.08\pm0.68$&SF&$\times$&-&$\times$\\
186717&17:02:01.824&31:53:01.341&$0.0332$&$9.54$&$6.99\pm0.68$&SF&$\times$&-&$\times$\\
189025&20:46:38.107&00:20:21.688&$0.0126$&$9.73$&$7.22\pm0.68$&LINER&$\times$&-&$\times$\\
208212&14:07:02.979&50:43:14.611&$0.0072$&$8.34$&$5.51\pm0.68$&Composite&$\times$&-&$\times$\\
215346&01:15:37.856&00:12:58.114&$0.0453$&$9.49$&$6.93\pm0.68$&SF&$\times$&-&$\times$\\
234033&09:54:38.791&40:32:04.395&$0.0673$&$9.52$&$6.97\pm0.68$&Seyfert&\checkmark&6.657&$\times$\\
235055&11:01:16.027&10:16:17.924&$0.0358$&$9.47$&$6.9\pm0.68$&SF&$\times$&-&$\times$\\
239541&12:33:07.855&10:14:56.145&$0.0531$&$9.61$&$7.08\pm0.68$&SF&$\times$&-&$\times$\\
241996&10:29:28.222&09:50:01.145&$0.0224$&$9.13$&$6.48\pm0.68$&SF&$\times$&-&$\times$\\
249598&14:26:31.705&44:51:42.018&$0.0079$&$9.04$&$6.36\pm0.68$&SF&$\times$&-&$\times$\\
250824&08:07:00.613&05:37:37.137&$0.052$&$9.47$&$6.9\pm0.68$&SF&$\times$&-&$\times$\\
266623&14:15:45.723&40:06:19.138&$0.0202$&$9.73$&$7.22\pm0.68$&SF&$\times$&-&$\times$\\
267136&14:42:17.705&40:02:006.81&$0.0266$&$8.87$&$6.16\pm0.68$&SF&$\times$&-&$\times$\\
273124&11:59:05.594&45:49:13.767&$0.024$&$9.56$&$7.01\pm0.68$&Composite&$\times$&-&$\times$\\
278949&15:36:19.446&30:40:056.33&$0.0059$&$8.36$&$5.52\pm0.68$&SF&$\times$&-&$\times$\\
280953&16:11:14.252&25:11:56.413&$0.0475$&$9.61$&$7.08\pm0.68$&SF&$\times$&-&$\times$\\
284032&16:02:50.886&32:08:41.391&$0.0141$&$9.49$&$6.92\pm0.68$&Composite&$\times$&-&$\times$\\
286477&15:48:36.238&36:13:42.486&$0.0396$&$9.46$&$6.89\pm0.68$&SF&$\times$&-&$\times$\\
295794&11:36:57.683&41:13:18.507&$0.0715$&$9.75$&$7.25\pm0.68$&Seyfert&\checkmark&6.51&$\times$\\
301231&13:18:29.733&45:17:01.821&$0.0358$&$9.67$&$7.15\pm0.68$&Composite&$\times$&-&$\times$\\
301767&13:35:42.788&45:55:46.505&$0.0048$&$8.95$&$6.26\pm0.68$&SF&$\times$&-&\checkmark\\
315748&08:02:13.476&22:26:02.457&$0.03$&$9.28$&$6.66\pm0.68$&SF&$\times$&-&$\times$\\
316044&08:19:50.791&24:47:16.158&$0.0256$&$9.65$&$7.12\pm0.68$&SF&$\times$&-&$\times$\\
319190&09:54:57.968&36:05:50.942&$0.0488$&$9.39$&$6.8\pm0.68$&SF&$\times$&-&$\times$\\
321176&10:43:26.474&11:05:24.227&$0.0475$&$9.64$&$7.12\pm0.68$&Seyfert&\checkmark&8.163&\checkmark\\
323877&11:46:04.072&11:34:52.566&$0.0099$&$9.18$&$6.54\pm0.68$&SF&$\times$&-&$\times$\\
323951&11:48:32.458&12:42:19.067&$0.0146$&$9.52$&$6.96\pm0.68$&SF&$\times$&-&$\times$\\
340153&14:38:58.286&47:18:28.497&$0.027$&$9.66$&$7.14\pm0.68$&SF&$\times$&-&$\times$\\
346014&13:09:15.702&13:23:08.333&$0.0274$&$9.23$&$6.61\pm0.68$&Composite&$\times$&-&$\times$\\
354455&15:17:39.678&11:34:59.736&$0.0652$&$9.5$&$6.94\pm0.68$&Composite&$\times$&-&$\times$\\
355963&15:43:58.996&08:07:34.691&$0.0409$&$9.62$&$7.09\pm0.68$&SF&$\times$&-&$\times$\\
358934&07:44:02.155&45:30:20.646&$0.0538$&$9.32$&$6.72\pm0.68$&SF&$\times$&-&$\times$\\
361340&10:07:10.915&12:39:06.079&$0.0093$&$8.47$&$5.67\pm0.68$&SF&$\times$&-&$\times$\\
369885&13:07:33.823&14:00:31.128&$0.0549$&$9.72$&$7.21\pm0.68$&SF&$\times$&-&$\times$\\
369924&13:07:17.443&13:38:47.968&$0.0268$&$9.66$&$7.14\pm0.68$&Composite&$\times$&-&$\times$\\
370005&13:06:056.11&14:48:26.805&$0.0032$&$8.1$&$5.21\pm0.68$&SF&$\times$&-&$\times$\\
370204&13:17:40.461&13:56:24.681&$0.0283$&$9.01$&$6.34\pm0.68$&SF&$\times$&-&$\times$\\
376082&12:51:55.683&08:52:41.595&$0.0038$&$8.45$&$5.63\pm0.68$&SF&$\times$&-&$\times$\\
376210&12:51:47.278&09:51:34.461&$0.0305$&$9.27$&$6.66\pm0.68$&SF&$\times$&-&$\times$\\
376231&12:54:07.922&10:00:23.692&$0.037$&$9.51$&$6.95\pm0.68$&Composite&$\times$&-&$\times$\\
376574&12:56:53.446&08:09:41.682&$0.0086$&$8.75$&$6.0\pm0.68$&SF&$\times$&-&$\times$\\
376622&12:58:21.864&08:32:29.792&$0.0289$&$9.56$&$7.01\pm0.68$&SF&$\times$&-&$\times$\\
377912&13:08:28.602&10:40:08.594&$0.0246$&$9.68$&$7.16\pm0.68$&SF&$\times$&-&$\times$\\
385012&15:12:02.597&06:55:53.926&$0.0465$&$9.73$&$7.23\pm0.68$&SF&$\times$&-&$\times$\\
385214&15:06:52.734&08:10:23.682&$0.0396$&$9.24$&$6.62\pm0.68$&SF&$\times$&-&$\times$\\
386591&15:26:37.364&06:59:41.726&$0.0384$&$9.39$&$6.8\pm0.68$&Composite&\checkmark&5.485&$\times$\\
389406&14:25:52.842&06:32:15.001&$0.024$&$9.21$&$6.57\pm0.68$&SF&$\times$&-&$\times$\\
391212&15:04:16.147&05:49:21.806&$0.0301$&$9.69$&$7.18\pm0.68$&SF&$\times$&-&$\times$\\
393092&16:03:43.582&04:32:35.996&$0.0444$&$9.69$&$7.17\pm0.68$&SF&$\times$&-&$\times$\\
401713&08:02:53.694&55:17:55.692&$0.03$&$9.41$&$6.83\pm0.68$&SF&$\times$&-&$\times$\\
404064&09:34:38.578&67:22:29.017&$0.023$&$8.89$&$6.18\pm0.68$&SF&$\times$&-&$\times$\\
412056&09:46:23.749&34:16:52.073&$0.0755$&$9.01$&$6.34\pm0.68$&SF&$\times$&-&$\times$\\
418509&13:34:05.899&27:33:40.195&$0.0293$&$9.5$&$6.94\pm0.68$&Composite&$\times$&-&$\times$\\
418517&13:35:26.025&28:10:12.928&$0.0633$&$9.63$&$7.1\pm0.68$&SF&$\times$&-&$\times$\\
418760&13:35:51.276&29:12:49.882&$0.0212$&$9.17$&$6.53\pm0.68$&SF&$\times$&-&$\times$\\
418763&13:35:35.715&29:13:03.022&$0.0028$&$8.19$&$5.32\pm0.68$&SF&$\times$&-&$\times$\\
424989&13:21:18.974&30:04:23.853&$0.0912$&$9.66$&$7.14\pm0.68$&SF&\checkmark&6.436&\checkmark\\
431557&13:42:05.443&32:01:043.28&$0.0156$&$9.46$&$6.89\pm0.68$&SF&$\times$&-&$\times$\\
437426&09:09:34.371&25:13:022.76&$0.0078$&$8.98$&$6.29\pm0.68$&Composite&$\times$&-&$\times$\\
437561&09:17:17.051&26:26:59.206&$0.0247$&$9.56$&$7.01\pm0.68$&SF&$\times$&-&$\times$\\
438165&12:03:25.676&33:08:46.147&$0.0349$&$9.1$&$6.45\pm0.68$&SF&$\times$&-&$\times$\\
439902&13:39:57.611&30:42:54.223&$0.042$&$9.65$&$7.12\pm0.68$&SF&$\times$&-&$\times$\\
439931&13:41:13.598&30:23:27.794&$0.04$&$9.47$&$6.9\pm0.68$&SF&$\times$&-&$\times$\\
439935&13:42:43.255&30:43:39.805&$0.0357$&$9.73$&$7.22\pm0.68$&Composite&$\times$&-&$\times$\\
439946&13:41:50.188&30:33:57.114&$0.0264$&$9.11$&$6.46\pm0.68$&SF&$\times$&-&\checkmark\\
445385&13:28:24.676&31:09:37.833&$0.0164$&$8.63$&$5.86\pm0.68$&-&$\times$&-&$\times$\\
451469&14:14:05.019&26:33:36.804&$0.0358$&$9.66$&$7.14\pm0.68$&SF&\checkmark&6.303&\checkmark\\
453882&14:28:30.841&27:15:57.271&$0.013$&$9.35$&$6.76\pm0.68$&SF&$\times$&-&$\times$\\
457334&14:47:17.407&24:30:19.296&$0.0424$&$9.46$&$6.89\pm0.68$&SF&$\times$&-&$\times$\\
465843&15:46:53.324&17:52:21.579&$0.0113$&$8.86$&$6.15\pm0.68$&SF&$\times$&-&$\times$\\
475108&11:30:26.846&30:27:52.084&$0.0589$&$9.66$&$7.14\pm0.68$&Composite&$\times$&-&$\times$\\
475368&11:28:28.072&27:54:07.612&$0.068$&$9.62$&$7.09\pm0.68$&Composite&$\times$&-&$\times$\\
477209&11:56:35.312&28:29:25.256&$0.0117$&$9.28$&$6.67\pm0.68$&Composite&$\times$&-&$\times$\\
480677&12:29:03.509&29:46:45.995&$0.0814$&$9.71$&$7.2\pm0.68$&SF&\checkmark&-&\checkmark\\
481291&12:36:012.26&26:45:22.078&$0.0246$&$9.71$&$7.2\pm0.68$&SF&$\times$&-&$\times$\\
484296&13:29:39.134&25:47:50.822&$0.0247$&$9.13$&$6.48\pm0.68$&SF&$\times$&-&$\times$\\
485578&08:04:18.001&15:20:038.47&$0.0393$&$9.73$&$7.22\pm0.68$&Seyfert&$\times$&-&$\times$\\
487099&08:25:21.331&15:17:023.31&$0.0329$&$9.49$&$6.93\pm0.68$&SF&$\times$&-&$\times$\\
487207&08:20:13.444&16:12:41.858&$0.0442$&$9.56$&$7.01\pm0.68$&SF&$\times$&-&$\times$\\
488733&08:41:028.47&16:16:43.165&$0.0731$&$9.72$&$7.21\pm0.68$&SF&$\times$&-&$\times$\\
501377&10:32:06.026&22:59:21.809&$0.0585$&$8.97$&$6.29\pm0.68$&Composite&$\times$&-&$\times$\\
502687&10:05:0023.1&19:16:18.857&$0.0128$&$8.73$&$5.98\pm0.68$&SF&$\times$&-&$\times$\\
508645&09:02:50.471&14:14:08.301&$0.0506$&$9.48$&$6.92\pm0.68$&-&$\times$&-&\checkmark\\
522620&11:42:50.986&20:26:31.655&$0.019$&$9.61$&$7.08\pm0.68$&SF&$\times$&-&$\times$\\
538951&12:32:36.165&18:01:23.085&$0.0028$&$8.67$&$5.91\pm0.68$&SF&$\times$&-&$\times$\\
540489&13:13:54.174&16:43:39.707&$0.0224$&$9.5$&$6.94\pm0.68$&SF&$\times$&-&$\times$\\
540494&13:13:40.008&16:39:27.698&$0.0219$&$8.98$&$6.3\pm0.68$&SF&$\times$&-&$\times$\\
548558&13:13:050.79&23:15:17.935&$0.0116$&$8.97$&$6.28\pm0.68$&SF&$\times$&-&$\times$\\
548726&13:23:016.62&21:18:005.97&$0.0224$&$9.61$&$7.08\pm0.68$&Composite&$\times$&-&$\times$\\
548780&13:20:42.199&20:54:37.122&$0.0093$&$8.44$&$5.63\pm0.68$&SF&$\times$&-&$\times$\\
552294&13:27:59.639&23:35:50.831&$0.0432$&$9.32$&$6.72\pm0.68$&SF&$\times$&-&$\times$\\
558933&14:09:28.215&17:39:43.879&$0.0181$&$9.67$&$7.15\pm0.68$&SF&$\times$&-&$\times$\\
563981&14:13:13.548&20:25:25.781&$0.0168$&$9.52$&$6.97\pm0.68$&SF&$\times$&-&$\times$\\
571135&15:10:24.257&19:23:47.709&$0.075$&$9.56$&$7.01\pm0.68$&SF&$\times$&-&$\times$\\
572805&12:25:18.255&05:44:31.206&$0.0038$&$9.11$&$6.46\pm0.68$&SF&$\times$&-&$\times$\\
576634&15:40:29.285&00:54:37.064&$0.0117$&$8.59$&$5.81\pm0.68$&SF&$\times$&-&$\times$\\
583865&00:56:33.363&00:05:010.27&$0.0793$&$9.45$&$6.88\pm0.68$&Seyfert&\checkmark&6.131&$\times$\\
592732&09:31:018.44&02:46:052.34&$0.1148$&$8.83$&$6.11\pm0.68$&Composite&$\times$&-&$\times$\\
593159&11:04:16.043&05:16:30.756&$0.1166$&$9.33$&$6.73\pm0.68$&Composite&$\times$&-&$\times$\\
600245&11:51:42.585&06:51:30.529&$0.102$&$9.18$&$6.55\pm0.68$&Composite&$\times$&-&$\times$\\
612283&12:05:03.537&45:51:51.056&$0.0654$&$7.52$&$4.48\pm0.68$&SF&$\times$&-&$\times$\\
627704&12:03:49.742&29:42:56.153&$0.0105$&$9.46$&$6.89\pm0.68$&SF&$\times$&-&$\times$\\
638093&23:32:45.037&00:58:45.886&$0.0238$&$9.45$&$6.88\pm0.68$&SF&$\times$&-&$\times$\\
648447&09:24:039.36&17:39:47.977&$0.0136$&$9.6$&$7.06\pm0.68$&Composite&$\times$&-&$\times$\\
661360&11:33:23.454&55:04:20.733&$0.0079$&$8.79$&$6.06\pm0.68$&SF&$\times$&-&\checkmark\\
676393&13:36:20.429&28:47:50.676&$0.0268$&$9.51$&$6.95\pm0.68$&Composite&$\times$&-&$\times$\\
677377&13:46:22.893&29:19:02.782&$0.0546$&$9.66$&$7.14\pm0.68$&SF&$\times$&-&$\times$\\
679695&14:07:24.993&55:06:10.896&$0.0046$&$9.22$&$6.6\pm0.68$&Composite&$\times$&-&$\times$\\
681375&14:27:09.515&30:56:53.601&$0.0137$&$9.59$&$7.06\pm0.68$&Composite&$\times$&-&$\times$\\
681956&14:34:037.05&59:20:018.29&$0.0064$&$9.03$&$6.35\pm0.68$&SF&$\times$&-&$\times$\\
683356&14:48:27.022&31:47:28.383&$0.0096$&$9.34$&$6.74\pm0.68$&SF&$\times$&-&$\times$\\
693966&12:59:20.083&09:11:13.787&$0.0279$&$9.66$&$7.14\pm0.68$&Composite&$\times$&-&$\times$\\
694183&13:13:24.659&08:02:24.125&$0.0237$&$9.71$&$7.2\pm0.68$&Composite&$\times$&-&$\times$\\
\enddata
\vspace{0.1cm}
\tablecomments{Properties of the 148 AGN candidates with significant and correlated W1 and W2 variability found in forward modeled \textit{WISE} light curves. The IDs, positions, redshifts and host galaxy stellar masses are those from the NSA catalog version \texttt{v$_{1\textunderscore0\textunderscore1}$}. In the  log$_{10}$M$_{BH}$ column we show the estimated black hole mass based on the $M_*-M_{BH}$ from \citet{Schutte2019TheNuclei} which has a scatter of 0.68 dex. The presence of Balmer broad lines is indicated in the BLR column. Virial masses were calculated for broad line AGN by \citet{Ho2015ASPECTRA} based on the width of the H$\alpha$ broad lines. \textbf{*}The estimated mass provided for NSA64525 by the NASA-Sloan Atlas ($10^{5.12}M\odot$) is inconsistent with the stellar dispersion velocity of $193\pm8$ km/s measured by the SDSS spectroscopic pipeline. This $\sigma_*$ is more characteristic of a host galaxy with mass $\sim10^{9.5}M_{\odot}$ and a BH with mass $\sim10^7M_{\odot}$. In the last column, we indicate whether the WISE candidate's ZTF light curve also met the criteria for optical variability.}
\end{deluxetable*}
\bibliography{main.bib}{}
\bibliographystyle{aasjournal}

\end{document}